\documentclass[aps,prl,twocolumn,superscriptaddress,showpacs,showkeys,amsfonts,amssymb,amsmath]{revtex4-1}
\usepackage{graphicx,color}
\usepackage{float}
\usepackage{amsmath}
\usepackage{xspace}
\usepackage{hyperref}
\usepackage[T1]{fontenc}
\newcommand{\degree}{\ensuremath{^\circ}\xspace}
\newcommand{\SxCO}{Sr$_{14-x}$Ca$_x$Cu$_{24}$O$_{41}$\xspace}
\newcommand{\SCO}{Sr$_{14}$Cu$_{24}$O$_{41}$\xspace}

\newcommand{\Tco}{$T_{\mathrm{co}}$\xspace}
\newcommand{\wvnmbr}{cm$^{-1}$\xspace}

\begin{document}


\title{Effects of Ca substitution on quasi-acoustic sliding modes in Sr$_{14-x}$Ca$_x$Cu$_{24}$O$_{41}$}


\author{E. Constable}
\email[]{evan.constable@tuwien.ac.at}
\affiliation{Institute of Solid State Physics, Vienna University of Technology, 1040 Vienna, Austria}
\affiliation{Institut N\'eel, CNRS and Universit\'{e} Grenoble Alpes, 38042 Grenoble, France}
\affiliation{Institute for Superconducting and Electronic Materials, University of Wollongong, Wollongong, NSW 2522, Australia}

\author {A. D. Squires}
\author{J. Horvat}
\author {R. A. Lewis}
\affiliation{Institute for Superconducting and Electronic Materials, University of Wollongong, Wollongong, NSW 2522, Australia}
\author{D. Appadoo}
\author{R. Plathe}
\affiliation{Australian Synchrotron, Australian Nuclear Science and Technology Organisation, 800 Blackburn Rd Clayton, VIC 3168, Australia}
\author {P. Roy}
\author {J.-B. Brubach}
\affiliation{Synchrotron SOLEIL, L'Orme des Merisiers, Saint-Aubin, BP 48, F-91192 Gif-sur-Yvette Cedex, France}
\author {S. deBrion}
\affiliation{Institut N\'eel, CNRS and Universit\'{e} Grenoble Alpes, 38042 Grenoble, France}
\author{A. Pimenov}\affiliation{Institute of Solid State Physics, Vienna University of Technology, 1040 Vienna, Austria}
\author {G. Deng}
\email[]{guochu.deng@ansto.gov.au}
\affiliation{Australian Centre for Neutron Scattering, Australian Nuclear Science and Technology Organisation, New Illawarra Rd, Lucas Heights, Sydney, NSW 2234, Australia}


\date{\today}

\begin{abstract}
The low energy lattice dynamics of the quasi-periodic spin-ladder cuprate \SxCO are investigated using terahertz frequency synchrotron radiation.
A high density of low-lying optical excitations are present in the 1-3 THz energy range, while at least two highly absorbing excitations stemming from rigid acoustic oscillations of the incommensurate chain and ladder sublattices, are observed at sub-terahertz frequencies.
The effects of Ca substitution on the sub-terahertz quasi-acoustic sliding mode gaps is investigated using coherent synchrotron radiation.
Analysis of the results suggest increasing substitution of Sr for Ca is accompanied by a transfer of spectral weight between sliding modes associated with different chain-ladder dynamics. The observation is consistent with a transfer of hole charges from the chains to the ladders and modification of the sublattice dimensions following Ca substitution.  
The results are discussed in context to the significance of low-lying vibrational dynamics and electron-phonon coupling in the superconducting state of certain quasi-periodic systems.

\end{abstract}


\maketitle

\section{Introduction \label{sec:intro}}

\begin{figure}
\centering
\includegraphics{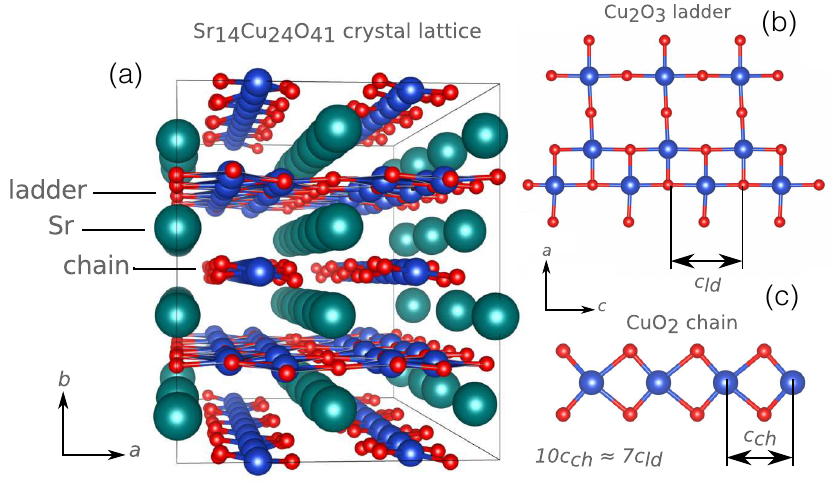}
\caption{(a) Super-cell structure of \SCO with $a$ = 11.47~\AA{}, $b$ = 13.41~\AA{} and $c$ = 27.5~\AA{} lattice parameters. (b) Section of Cu$_2$O$_3$ two-leg ladder sublattice with $c_{\mathrm{ld}}$ = 3.95~\AA{}. (c) Section of CuO$_2$ chain sublattice with $c_{\mathrm{ch}}$ = 2.75~\AA{}. }
\label{fig:crystLattice}
\end{figure}

The significance of neighboring orders and their collective excitations in the thermodynamic landscape of high temperature superconductors (HTS) is an ongoing question within condensed matter physics. The simple notion of whether or not such orders ultimately enhance or impede superconductivity is examined in numerous HTS systems \cite{Johnston2010,Gallais2016,Manzeli2017}. Recently, the implication of quasi-periodic lattice order with its associated low energy vibrational modes has also received interest in attempts to understand examples of strong-coupling superconductivity \cite{Klintberg2012,Tompsett2014,Brown2018,Khasanov2018,Santoro2018,Cheung2018}.
In this context, the quasi-periodic spin-ladder cuprate \SCO (SCO) presents a unique environment to explore these ideas.
Built from alternating layers of Sr$_2$ planes, two-leg Cu$_2$O$_3$ ladders and CuO$_2$ chains (see Fig.~\ref{fig:crystLattice}), SCO forms a quasi-periodic superlattice due to an incommensurate free arrangement of the chain and Sr-ladder sublattices along the $c$ direction, with $c$ = 27.5~\AA{} $\simeq$ 10$c_{\mathrm{ch}}$$\simeq$ 7$c_{\mathrm{ld}}$ \cite{Mccarron1988}, where $c_{\mathrm{ch}}$ and $c_{\mathrm{ld}}$ are the $c$ lattice parameters of the chain and ladder sublattices, respectively.
A nominal valence of Cu$^{2.6+}$ for each CuO$_2$ site intrinsically dopes the chains with $\sim$6 hole charge carries per unit formula \cite{Nucker2000}. 
This, in combination with the quasi-periodic lattice, facilitates a number of exotic low-dimensional charge and S = 1/2 quantum magnetic properties including anisotropic non-linear charge transport \cite{Akimitsu1996,Kato1996_2,Adachi1998}, charge density wave (CDW) charge order \cite{Cox1998,Blumberg2002,Gorshunov2002,Choi2006}, hole crystallization \cite{Abbamonte2004}, spin gapped antiferromagnetic dimerization \cite{Eccleston1996,Matsuda1996,Matsuda1996_2,Eccleston1998,Matsuda1999}, and long-range quantum coherence \cite{Lorenzo2010}.

Substitution of Sr with Ca in compounds of \SxCO (SCCO) acts to transfer holes from the chains to the ladders, providing control over the unique physics expressed in each sublattice and hence tunability over the landscape of electronic orders \cite{Vuletic2003,Nucker2000,Deng2011_2,Huang2013,Deng2013,Bag2018}. Perhaps the most significant consequence being the emergence of superconductivity below 12 K for a Ca concentration of  $x=13.6$ and hydrostatic pressure of 3 GPa \cite{Uehara1996}.
The occurrence of superconductivity in SCCO is noted as unique amongst the HTS cuprates because the lattice maintains its one-dimensional nature at high pressure and lacks the familiar CuO$_4$ square plackets \cite{Isobe1998}. As such, the nature of superconductivity in SCCO remains poorly understood.
Remarkably, recent spectroscopic experiments on the order of 1 meV have revealed the presence of additional acoustic vibrational modes supported by a rigid sliding oscillation of the incommensurate chain and ladder sublattices \cite{Thorsmolle2012,Chen2016}. Theoretically predicted over 30 years earlier \cite{Theodorou1978,Theodorou1980}, these quasi-acoustic sliding modes invoke similarities to the low-frequency vibrational modes of other quasi-periodic superconductors \cite{Tompsett2014,Santoro2018,Cheung2018,Brown2018} and therefore warrant further attention.

Spectral studies of low energy dynamics has proven to be an effective probe of the interplay between the different electronic orders in SCCO.
However, while considerable effort has been made to understand the effects of Ca substitution on the charge  order \cite{Ruzicka1998,Vuletic2003,Nucker2000} and spin-gap \cite{Motoyama1997,Deng2013,Bag2018,Deng2018} properties of SCCO, little attention has been given to the effects on the quasi-acoustic sliding modes.
To address this we have attempted a spectroscopic study on a series of SCCO samples using a combination of synchrotron and laboratory-based terahertz (THz) and far-infrared (FIR) techniques.
In combination, the different techniques provide a picture of the low energy lattice dynamics in relation to changes in temperature and Ca concentration.
We observe a number of highly anisotropic optical excitations below 3 THz including strong absorption bands in the sub-THz energy range of SCCO consistent with previous works \cite{Gorshunov2002, Vuletic2003,Thorsmolle2012,Chen2016}. Our results show a transfer of spectral weight between sliding mode excitations associated with slow in-phase and fast out-of-phase chain-ladder sliding dynamics. We note that it is within the ladders that the superconducting channels are expected and discuss the implications of our results in this context.


\section{Experimental details\label{sec:methds}}

Single crystals of SCCO with $x$ = 0, 3, 7 and 11 were grown by the floating zone method and prepared as wafers cut normal to the $b$ axis. Sample dimensions were approximately 5$\times$5$\times$1 mm$^3$. Details of the sample characterization can be found in reference \cite{Deng2011}.
Transmission measurements of the $x$ = 0 sample in the 0.6--3.6~THz 
spectral range from 5.6--300~K were performed at the Australian Synchrotron far-infrared beamline. 
Spectra were collected using a Bruker IFS125HR FTIR Michelson interferometer with a 75~$\mu$m mylar beam splitter and 4.2~K Si Bolometer detector at 0.5 \wvnmbr resolution for polarized $E||c$ and $E||a$ configurations.
A 2.5~mm aperture copper mount was used as a reference.
Complementary spectra in the same energy range were also collected for the $x$ = 0 and $x$ = 3 samples using a modified polytec FIR 25 Fourier transform interferometer with mercury lamp source, interfaced to the low-temperature insert of a He cooled 7-T Oxford split-coil superconducting magnet. To test the magnetic activity of observed excitations, experiments were performed at 5 K from 0 to 5 T. However, we observed no magnetic field effects and therefore only 0 T measurements are presented here.
Transmission spectra of the $x$ = 0, 3, 7 and 11 samples in the 0.2--1.2~THz (6.7--40~\wvnmbr) spectral range were performed at the SOLEIL synchrotron AILES beamline operating with coherent synchrotron radiation in a low-$\alpha$ configuration providing a high power, polarized radiation source in the sub-THz energy range \cite{Barros2013,Barros2015}.
The samples were mounted to an Attocube rotation stage thermally coupled to the cold head of a Cryomech PT 405 pulse tube cryostat, achieving sample temperatures from 16--300~K.
Spectra were collected using a Bruker IFS125 FTIR Michelson interferometer with a 125~$\mu$m mylar beam splitter and 1.6~K Si Bolometer detector at 0.5 \wvnmbr resolution. Experimental limitations prevented measurement of a conventional vacuum reference. Therefore, quantitative extraction of sample attenuation was determined by analyzing the rotationally dependent response while assuming a flat contribution for $E||a$. Details of this analysis process are covered in the supplementary material \cite{SupMat}.

\section{Results \label{sec:results}}

\begin{figure*}
\centering
\includegraphics{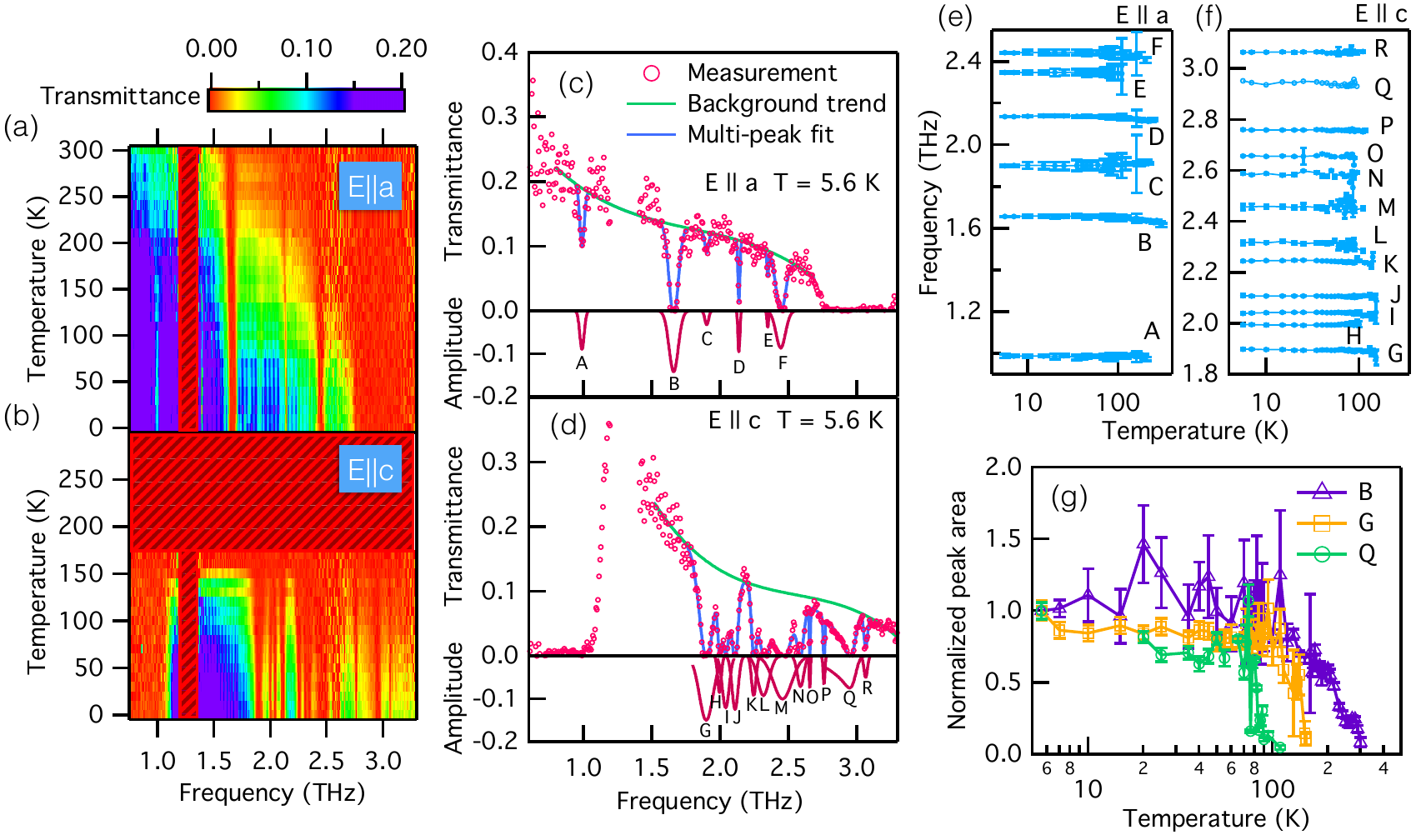}
\caption{Temperature dependence of SCO transmission between 0.7 and 3.3 THz for $E||a$ (a) and $E||c$ (b) determined by a ratio with a vacuum reference. Opaque spectra for $E||c$ above 170 K were omitted and are represented by a hatched rectangle. Low signal intensity due to a beamsplitter minimum results in noise artifacts centered on 1.29 THz and are also omitted. Multi-peak fits of the spectra at 5.6 K are shown for $E||a$ (c) and $E||c$ (d). Peaks were fit using Gaussian profiles and a background trend approximated by a 3rd order polynomial. Temperature dependence of fitted peak positions for $E||a$ (e) and $E||c$ (b). (g) Temperature dependence of fitted peak intensity for selected peaks B, G and Q determined by peak area normalized to the base temperature (5.6 K) value.}
\label{fig:AS_specs}
\end{figure*}

The general anisotropic and temperature dependent behavior for the conductivity and lattice dynamics of SCO in the far-infrared energy range is captured in the transmission spectra of Fig.~\ref{fig:AS_specs}. 
As seen in Figs.~\ref{fig:AS_specs} (a) and (c), the sample is semi-insulating at low frequency when the electric field is perpendicular to the chains and ladders ($E||a$), transmitting $\sim$10\% below 1 THz at room temperature. As the temperature is lowered, the sample transmission increases steadily as a broad absorption continuum shifts to higher frequencies, stabilizing at $\sim$2.75~THz below $\sim$100~K. 
This behavior is consistent with reflectivity measurements indicating a suppressed optical conductivity due to a shifting plasma response \cite{Ruzicka1998}.
It is worth noting that the stabilization of this response below 100~K corresponds to the temperature of spin dimer formation observed in the magnetic susceptibility \cite{Matsuda1996_2,Matsuda1996}. It suggests a correlation between charge transport and magnetic degrees of freedom in the $a$ direction and marks a crossover from 2D to 1D charge dynamics.
On the other hand, when the electric field is parallel to the chains and ladders ($E||c$, Figs.~\ref{fig:AS_specs} (b) and (d)), the sample is non-transparent at high temperatures due to the metallic behavior along $c$. Upon cooling below ~$\sim$150 K, the transmission abruptly increases following insulating behavior as a result of a pinned charge density wave \cite{Osafune1999,Vuletic2003,Choi2006} in the charge ordered phase emerging at \Tco = 200 K \cite{Adachi1998,Cox1998}.

In both polarizations a number of sharp absorption features are seen, which exhibit abnormally low energy, high density and fine structure when compared to conventional optical phonon modes in other alkaline cuperates \cite{Homes1993,Gruninger1998,Room2004,Ortolani2006}. 
Multi-peak fitting and error analysis of the absorption features (Figs.\ref{fig:AS_specs} (c) and (d)) was performed using a Marquardt-Levenberg nonlinear least-squares fitting procedure implemented within the Wave Metrics Igor Pro software package.
Results of the temperature dependence for fitted peak positions labeled from A-R are shown in Figs.\ref{fig:AS_specs} (e) and (f), while the peak intensity of three representative features (B, G, Q) is shown in Fig.\ref{fig:AS_specs} (g).
Minimal broadening and frequency shifts are observed as a function of temperature. 
The peak intensity profiles reveal three temperature regimes associated with the different electronic orders of SCO. In the high temperature regime (200 - 300 K), the dependence of peak B shows a general increase in phonon intensity as the temperature is lowered. We attribute this to the decreasing optical conductivity that minimizes screening by mobile charges and enhances the observed phonon intensity.
Below $\sim$200 K, a similar effect is observed along the $c$ direction following the stabilization of the insulating charge ordered phase that is accompanied by the appearance of new modes such as peak G.
Below $\sim$100 K, SCO is characterized by the stabilization of short range magnetic spin dimmer order on the chain sublattice. 
As observed in similar quasi-one dimensional charge and spin-dimmer ordered systems such as the (TMTTF)2X charge-transfer salts, such processes are often accompanied by local charge redistribution that stabilize new vibrational modes \cite{Dumm2006,Pustogow2016}. 
The presence of new modes such as peak Q, and the perceived splitting of peaks I and H below $\sim$100 K could be attributed to similar behavior in SCO.
Overall, the observed lattice dynamics are consistent with the highly ordered super-cell structure of SCO,  supporting a large number of under-damped, weakly dispersive optical lattice modes and demonstrating a mixing between lattice and electronic order parameters over a broad temperature range \cite{Ruzicka1998,Popovic2000,Sugai2001,Chen2016}. We note that the particularly high density of low lying phonons observed for the $c$ direction are also likely complimented by zone folded modes due to the incommensurate chain-ladder relationship and charge order Peierls distortion that modulates the chain and ladder symmetries while maintaining the same overall crystal structure \cite{Isobe1998,Braden2004,Rusydi2008}.
The zone folding phenomenon is well documented in other low dimensional systems exhibiting charge ordering \cite{Eldridge1985,Homes1989,Damascelli2000,Gorshunov2013}. In SCO the effect has previously been noted by the appearance of new phonon-like Raman modes below \Tco, in the vicinity of 6--18\,THz \cite{Popovic2000}. 
In general however, due to the large and complex nature of the SCO lattice, a more precise assignment of individual lattice excitations in Fig.~\ref{fig:AS_specs} remains a challenge.

Finally, we note that measurements performed under magnetic field (not shown) revealed no discernible peak shifts except for a weak signal resulting from Zeeman splitting of the chain localized spin triplet excitations at 2.33 THz (peak E) and 2.63 THz in the $E||a$ configuration. Details of the spin triplet excitations from chain and ladder localized Zhang-Rice singlet states have been covered elsewhere \cite{Matsuda1996,Eccleston1996,Eccleston1998,Matsuda1999} with an analysis of their optical excitation given in Ref.~\cite{Huvonen2007}. A further analysis of peak fits in the spectral region of Fig.~\ref{fig:AS_specs} is addressed in Ref.~\cite{Constable2014}. More details of the temperature dependence for the peak fitting parameters are provided in \cite{SupMat}.

\begin{figure}
\centering
\includegraphics{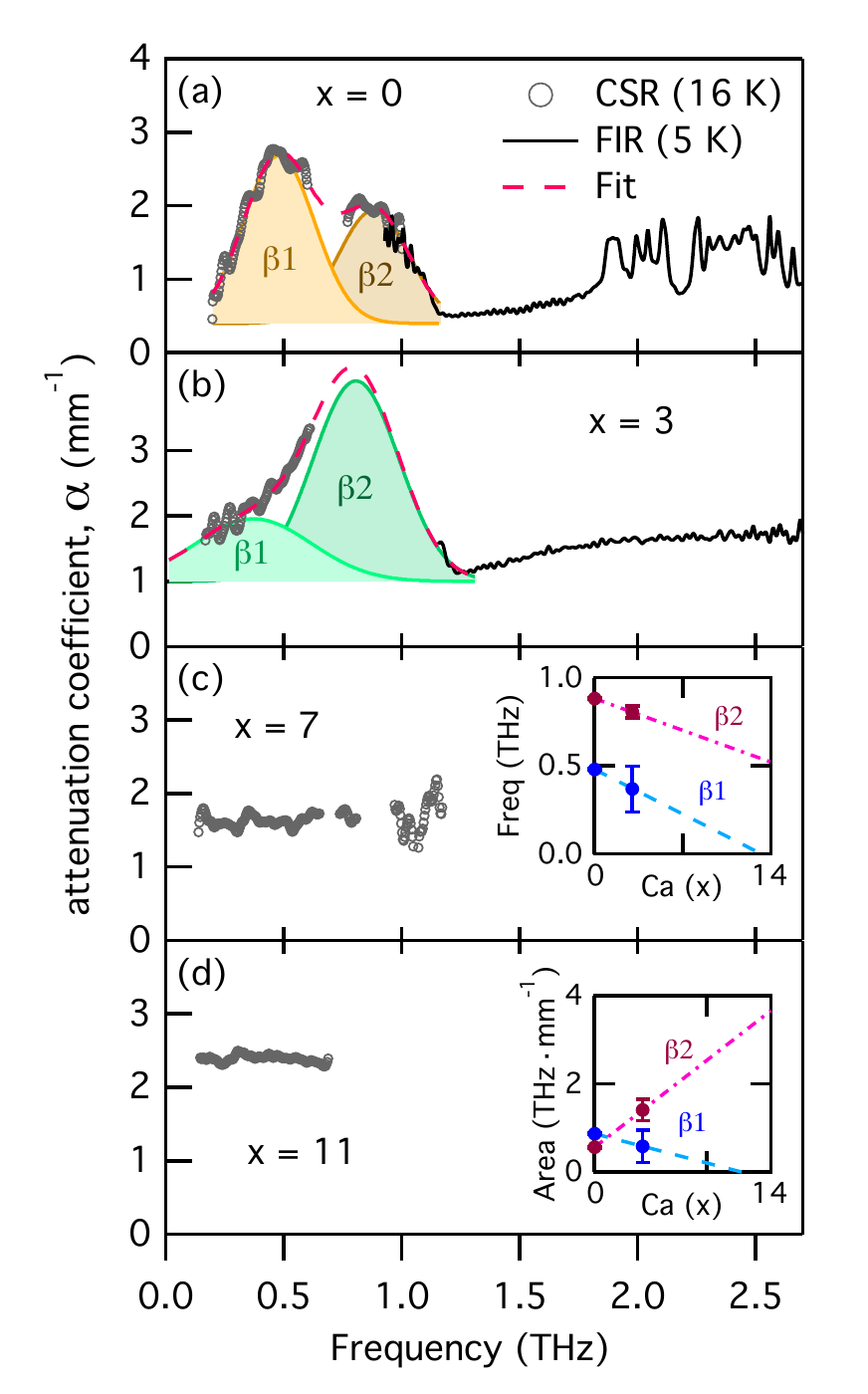}
\caption{Low frequency $E||c$ sample attenuation for SCCO with increasing Ca substitution. Peaks in the spectra correspond to resonant excitations in the sample. All displayed CSR data was measured at 16 K while the FIR data was measured at 5 K. Quantitative analysis of the CSR data was performed using the technique outlined in \cite{SupMat} with reference angle relative to $E||c$ of $\theta$ = 12\degree, 24\degree, 29\degree and 90\degree for the samples with Ca concentrations of $x$ = 0, 3, 7 and 11 respectively. The FIR data was analyzed using a conventional vacuum reference. Insets show approximate trends of the fitted peak position ((c) insert) and area ((d) insert) as a function of Ca substitution.}
\label{fig:Spec_CaDep}
\end{figure}

Apart from the high density of optical phonon modes beginning at $\sim$2 THz, a strong absorption band below $\sim$1 THz also dominates the spectra in the $E||c$ configuration (Figs.~\ref{fig:AS_specs} (b) and (d)).
The position, intensity, and anisotropy of the band suggests that it is related to gapped acoustic excitations stemming from ridged collective oscillations of the incommensurate chain and ladder sublattices. Such sliding modes are theoretically predicted in low dimensional incommensurate systems \cite{Theodorou1978,Theodorou1980} and were first identified in SCO by Thorsm\o{}lle \textit{et al.} \cite{Thorsmolle2012}.
The strong spectral intensity of these excitations renders the spectrum completely opaque to conventional THz transmission techniques and gives a flat reststrahlen band close to 1 in reflection. These limitations make precise optical measurement of the sliding mode gap difficult. In Ref.~\cite{Thorsmolle2012}, combined transmission and reflection techniques were used to give values of 0.25 THz and 0.38 THz for infrared active modes with different relative sliding motion.
Inelastic neutron scattering on SCO single crystals confirms the highly anisotropic acoustic dispersion of the sliding modes with a gap of 0.46 THz \cite{Chen2016}.
Together, these studies provide a convincing picture of the sub-THz excitations in SCO. 
However, the discrepancy in the observed gaps and their $\sim$50\% lower frequency compared to the absorption band beginning at $\sim$1 THz in Fig~\ref{fig:AS_specs} suggests that there are more details to be uncovered.
Moreover, a study of the effects of Ca doping on the properties of the sliding mode excitations is pertinent in relation to the role of low-frequency lattice modes in the superconducting state \cite{Brown2018}.
To address these points we performed THz transmission measurements on samples of SCCO with $x$ = 0, 3 ,7 and 11 using coherent synchrotron radiation.
The high flux at low frequency achievable with coherent synchrotron radiation allows measurement of strongly absorbing excitations in transmission geometry without complete attenuation of the signal at the resonant energy.
Taking advantage of the one dimensional anisotropic optical response of the samples at low temperature and the naturally polarized radiation of the low-$\alpha$ mode coherent synchrotron, spectra were analyzed by comparing changes in transmitted intensity as the sample is rotated away from a $E||c$ configuration. Following the details provided in \cite{SupMat}, and assuming a flat, sub-THz transmittance for $E||a$ of $T_a$~=~0.35 extrapolated from Fig.~\ref{fig:AS_specs}  for the $x$ = 0 sample, a quantitative measurement of the attenuation coefficient for $E||c$ is possible.
Similar analysis techniques have been implemented in probing highly absorbing excitations in other strongly anisotropic cuprates \cite{Pimenov2002}.
The results at 16 K are shown in Fig.~\ref{fig:Spec_CaDep} in combination with higher frequency measurements obtained by conventional far infrared methods at 5 K. The data is omitted between 0.6 and 0.77 THz due to noise artifacts at the beam splitter minimum.

Despite limitations of the rotational measurement technique and omitted data at the beam splitter minimum, a clear and broad resonant excitation band is observed centered around 0.6 THz for the $x$ = 0 sample in Fig.~\ref{fig:Spec_CaDep} (a). 
We attribute weak fringes on top of the excitation band to a phase mismatch between the plane parallel interference fringes for $E||c$ and $E||a$ configurations.
A nice agreement with the absorption band edge at $\sim$1 THz obtained by conventional FIR methods validates the quantitative analysis of the CSR data and provides a broadband picture of the low frequency optical response along $c$.
We find that the asymmetric sub-THz absorption band is fit best by two Gaussian profiles with energies of 0.48$\pm$0.01 and 0.88 $\pm$ 0.01 THz designated $\beta1$ and $\beta2$ respectively.
Data for the $x$ = 3 Ca substituted sample are shown in Fig.~\ref{fig:Spec_CaDep} (b). 
Here, a combination of the beam splitter minimum and strong sample absorption attenuates the signal between 0.62 and 1.14 THz.
Nevertheless, the combined results of the CSR and FIR techniques come together to show a consistent, quantitative picture of the low-frequency spectrum.

Firstly, we note the absence of the optical phonons above 1.5 THz that marks a distinction between the low-frequency optical lattice dynamics in the $x$ = 0 and $x$ = 3 compounds. 
This may be a result of hardening  due to increased lattice strain pushing the phonons above the detected spectral bandwidth.
It may also mark a suppression of zone folding behavior resulting from a different CDW modulation due to the expected transfer of holes from the chains to the ladders with increasing Ca substitution \cite{Nucker2000}.
While the effects of Ca substitution on the optical lattice dynamics appear dramatic, the effects on the sliding acoustic dynamics are subtler. 
As can be seen in Fig.~\ref{fig:Spec_CaDep} (b), the CSR data shows a rising absorption band from 0.2 to 0.6 THz. 
On the other hand, the FIR data shows an absorption band edge starting to form below 1.2 THz and suggests the presence of a resonant excitation below this frequency. 
While the attenuated spectrum between 0.62 and 1.14 THz obscures the complete spectral profile, we find a suitable fit to the available data is given by two Gaussian peak profiles with energies 0.37 $\pm$ 0.13 THz and 0.81 $\pm$ 0.03 THz as shown in Fig.~\ref{fig:Spec_CaDep} (b).
A resemblance to the sub-THz peaks in the $x$ = 0 sample is evidence that the two peaks in Fig.~\ref{fig:Spec_CaDep} (b) are the same gaped sliding modes designated $\beta1$ and $\beta2$ respectively. 
While we note that it is difficult to draw solid conclusions about the evolution of the spectral profile between the two samples in the absence of an intermediate measurement between $x$ = 0 and $x$ = 3 Ca concentrations, the analysis suggests there is a transfer of spectral weight from peak $\beta1$ to $\beta2$ with increasing Ca substitution.
The dependence of peak position and area on Ca substitution, including an extrapolated linear trend, is shown in the insets of Figs.~\ref{fig:Spec_CaDep} (c) and (d). 

Only low-frequency CSR data is available for the $x$ = 7 and 11 samples, shown in Figs. ~\ref{fig:Spec_CaDep} (c) and (d). 
In these two cases the sub-THz response is mostly flat. Based on the available literature \cite{Vuletic2003}, it is possible the $x$ = 7 sample is not completely stabilized in the charge ordered phase at 16 K. The $x$ = 11 sample should be either a poor insulator or completely metallic at 16 K.
Therefore it is likely that freely mobile charges on the chains and ladders in the $x$=7 and 11 samples act to screen resonance excitations in the $E||c$ configurations. This could then explain the absence of any resonant signatures of the gaped sliding modes in this energy range.



\begin{figure*}
\centering
\includegraphics{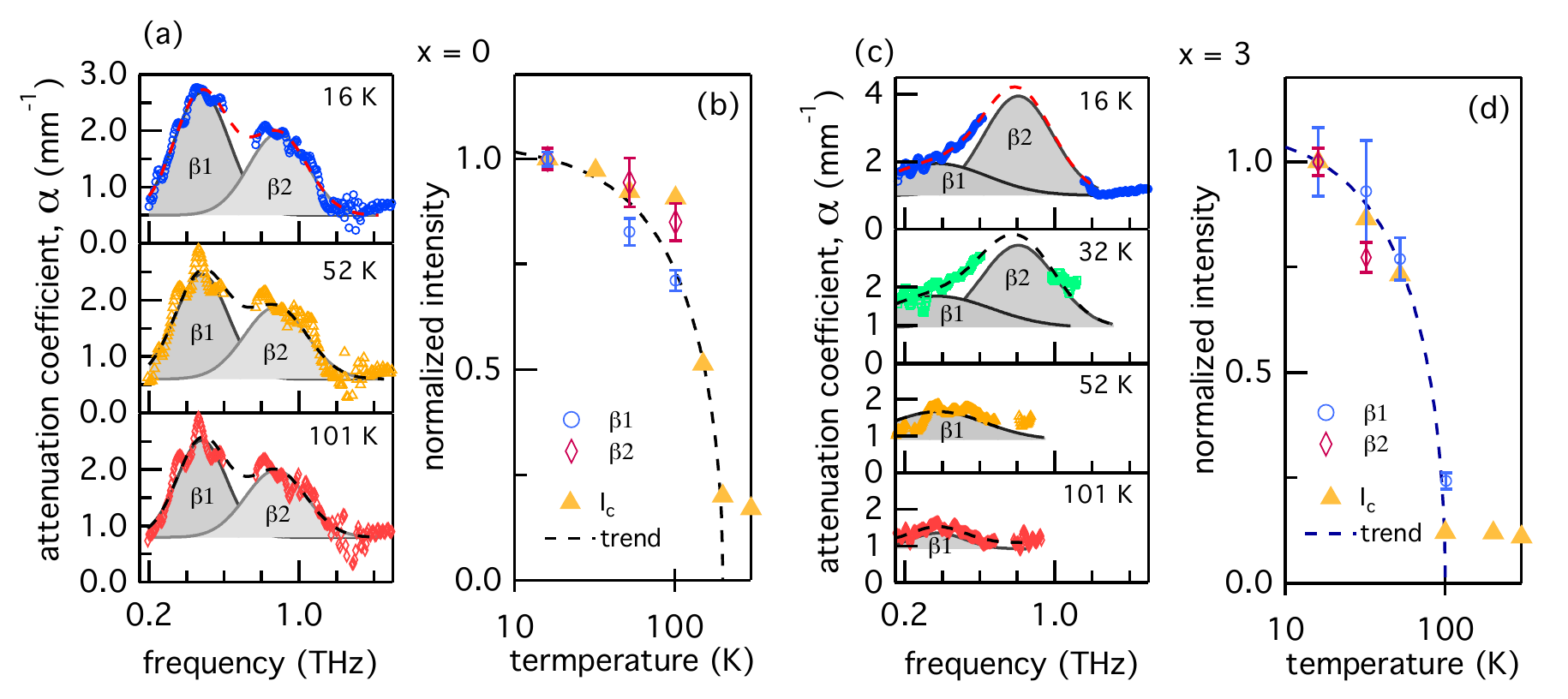}
\caption{Temperature dependent CSR spectra and fitted peak parameters of sliding mode gaps in SCCO for $x$ = 0 (a)-(b) and $x$ = 3 (c-d). Results above 0.9 THz are combined with the synchrotron spectra in Fig.~\ref{fig:AS_specs} for $x$ = 0 and with the FIR spectra at 5 K in  Fig.~\ref{fig:Spec_CaDep} (b) for $x$ = 3. The temperature dependence of broadband transmitted intensity between 0.1 and 1.2 THz for $E||c$ ($I_c$) along with a $(T_{\mathrm{CO}}-T)^{1/2}$ trend is plotted in (b) and (d). \Tco = 200 K is used for $x$ = 0, while \Tco = 100 K is used for $x$ = 3.}
\label{fig:Spec_TempDep}
\end{figure*}

We now turn our attention to the temperature dependence of the sliding mode excitations for the $x$ = 0 and $x$ = 3 samples shown in Fig.~\ref{fig:Spec_TempDep}.
As established in the spectra of Fig~\ref{fig:AS_specs}, as well as in the literature \cite{Ruzicka1998,Choi2006,Vuletic2003,Osafune1997,Schmidt2003}, the relationship between the charge and lattice dynamics in SCCO is highly temperature dependent. Moreover, this dependence shifts for different Ca concentrations and should be considered when comparing results for different samples at a single temperature.
As such, we would like to rule out any effects of temperature on the observed transfer of spectral weight between peaks $\beta1$ and $\beta2$ for $x$ = 0 and $x$ = 3 samples in Fig.~\ref{fig:Spec_CaDep}.
The results show negligible shifts in the sliding mode gap for $x$ = 0 as a function of temperature evidenced in both the CSR (Fig.~\ref{fig:Spec_CaDep}) and synchrotron (Fig~\ref{fig:AS_specs}) spectra.
Peak intensity determined by the fitted peak area for $\beta1$ and $\beta2$ as a function of temperature is shown in Fig.~\ref{fig:Spec_TempDep} (b) normalized to the value at 16 K. The fitting results are compared with normalized measurements of the average transmission ($I_{\mathrm{c}}$) in the 0.1 to 1.2 THz range reflecting the general change in optical conductivity.
A trend line is included of the form $(T_{\mathrm{CO}}-T)^{1/2}$ with \Tco = 200 K, approximating the charge ordered phase transition at 200 K for $x$ = 0.
Evidently, the excitation intensity closely follows the charge order temperature dependence.
As expected, the increasing conductivity of the sample at higher temperature will allow more effective screening of the sliding mode resonance reducing its spectral intensity.
A similar temperature dependence is found for the fitted peaks of the $x$ = 3 spectra shown in Figs.~\ref{fig:Spec_TempDep} (c)-(d). 
Here the charge order trend is shown using a critical temperature of \Tco = 100 K. 
Interestingly, we find that while peak $\beta1$ generally follows the charge ordering  trend, $\beta2$ decreases more rapidly appearing completely suppressed at higher temperatures.
Together, the results provide supporting evidence for the intrinsic nature of the observed transfer of spectral weight between $\beta1$ and $\beta2$ in the measurements of the $x$ = 0 and $x$ = 3 samples at 16 K.


\section{Discussion}
We point out that our observations of strongly absorbing sub-THz excitations in SCCO for $x$ = 0 and 3 are in close agreement with a previous study noting excitations at similar energies for the two compositions \cite{Gorshunov2002,Vuletic2003}. These resolution limited results were originally interpreted as density wave dynamics. However, subsequent optical \cite{Thorsmolle2012} and neutron scattering \cite{Chen2016} experiments have provided a more convincing picture, describing them  as gapped sliding or quasi-acoustic modes.
Such excitations should manifest in low dimensional systems with incommensurate sublattices such as in SCCO. This is because the ridged translation of one sublattice relative to the other leaves the total system energetically invariant due to their continuous symmetry relation \cite{Axe1982}. This effectively introduces additional acoustic modes on top of the conventional acoustic oscillations of the whole structure \cite{Theodorou1978}.
If the two incommensurate sublattices ($a$ and $b$) are oppositely charged, then analogous to electronic plasmon oscillations, the sliding modes will induce a net capacitance across the sample acting as a restoring force for the displacement. In a three-dimensional sample, the resorting electric field is independent of the distance of charge separation, sensitive only to the charge per unit area $\sigma$, and the residual permitivity of the lattice $\epsilon_{\mathrm{r}} = \epsilon_{0}\epsilon_{\infty}$ ($E=\sigma/\epsilon_{0}\epsilon_{\infty}$). Therefore, even at the long wavelength limit, a finite energy is required to excite such quasi-acoustic sliding modes,
with the zone center energy gap given by \cite{Theodorou1980},
\begin{equation}
\label{eq:Gap}
\omega^2 = \frac{n_aq_{a}^2}{\epsilon_{0}\epsilon_{\infty}m_{a}}\left(1 + \frac{n_b}{n_a}\frac{m_a}{m_b}\right),
\end{equation}
where  $n_{a,b}$, $q_{a,b}$ and $m_{a,b}$ are the incommensurate $a$ and $b$ sublattice unit cell density, charge and mass respectively. 

In the case of SCCO the CuO$_2$ chains feature a nominal charge of $q_{\mathrm{ch}}$ = (-2+0.6)e = -1.4e due to the localization of 6 residual holes over 10 chain units, while the Sr$_2$Cu$_2$O$_3$ Sr-ladder combination has a net charge of $q_{\mathrm{ld}}$ = +2e.
We can then estimate the sliding mode energy gap by considering a residual permitivity of $\epsilon_{\infty}$ = 15, consistent with reflectivity measurements available in the literature \cite{Ruzicka1998,Thorsmolle2012}, unit cell volumes of $V_{\mathrm{ch}}$ = 418 \AA$^3$, $V_{\mathrm{ld}}$ = 603 \AA$^3$\cite{Mccarron1988,Isobe1998} and masses of $m_{\mathrm{ch}}$ = (58+2$\times$16)$m_{\mathrm{p}}$, $m_{\mathrm{ld}}$ = (2$\times$76+2$\times$58+3$\times$16)$m_{\mathrm{p}}$, where $m_{\mathrm{p}}$ is the proton mass. Using these parameters with Equation \ref{eq:Gap} we find a sliding mode energy of 0.46 THz, both for modes involving chains sliding relative to static ladders and ladders sliding relative to static chains (see \cite{SupMat}). The result is in close agreement to our observation of peak $\beta1$ = 0.48 THz. In fact the small difference between the calculated and observed energies can be accounted for by a slight modification to the residual permitivity ($\epsilon_{\infty}\approx $ 14) within the uncertainty of typical reflectivity measurements, or by considering hole delocalization from the chains to the ladders (giving $q_{\mathrm{ch}}$ = -1.45e, $q_{\mathrm{ld}}$ = +2.1e) as could be expected in a $x$ = 0 SCCO sample with slight impurities \cite{Nucker2000}. It is also worth noting that the sensitivity of the sliding mode excitations to the permitivity and charge parameters as well as the noted variance in the hole concentrations of different samples \cite{Nucker2000} could explain the small frequency discrepancies between different observations in the literature \cite{Thorsmolle2012,Gorshunov2002,Vuletic2003,Chen2016}.
Moreover, the derivation of Equation~\ref{eq:Gap} assumes the case $k\gg 1/L$ where $k$ is the probing radiation wave vector and $L$ the sample length along $c$. While $k>1/L$ is valid for a 5 mm long sample and ~1 THz radiation, for shorter samples and longer wavelengths experiments approach $k = 1/L$, meaning sample dimensions may also play a role in the slightly different energy gaps observed. 

Applying the same analysis to peak $\beta2$ predicts chain and ladder sublattice charges on the order of $q$ = 2.6-3.8e, higher than the expected net valencies of the individual sublattices. Rather, we expect that this is evidence of a higher order harmonic in which the two sublattices slide out of phase relative to each other (see \cite{SupMat}), effectively increasing the charge separation and hence the strength of the restoring electric field. Considering Cu$_2$O$_3$ ladders decoupled from the Sr$_2$ ions, sliding out of phase with the CuO$_2$ chains and producing a combined charge displacement of $q$ = 3.4e we calculate a sliding mode energy gap of 0.88 THz in close agreement with the observation of $\beta2$. While this interpretation provides one description for a higher order sliding mode, we must also note that similar energy gaps are found by considering different scenarios in which lighter sublattices within the supercell structure move relative to each other. For example decoupling the chain motion from most of the ladder sublattice mass so that they are coupled only to the 180\degree bonded O$_2$ atoms in the ladders also gives an energy equivalent to $\beta2$. Moreover, the possibility that $\beta2$ represents a different sort of dynamic stemming from the charge ordering  cannot be completely ruled out with the experimental evidence available. 


Nevertheless, with a general understanding of the contributions to the sliding mode gap, we can now consider the different observations in Figs.~\ref{fig:Spec_CaDep} and \ref{fig:Spec_TempDep}. As established by optical \cite{Osada2000} and x-ray \cite{Nucker2000} techniques, Ca substitution in SCCO acts to reduce the distance between chain and ladder layers facilitating the transfer of holes from the chains to the ladders leaving the total number of holes roughly unchanged.
This will effectively increase the negative charge on the chains while increasing the positive charge on the ladders.
Moreover, Ca substitution acts to reduce the chain and ladder unit cell volumes as well as the Sr$_2$Cu$_2$O$_3$ sublattice mass \cite{Mccarron1988,Isobe1998}.
Each of these processes will have the net effect of increasing the energy gap for all sliding mode types.
In contrast, reflectivity measurements show an increase in the dielectric permitivity following Ca substitution \cite{Ruzicka1998} that would facilitate softening of the sliding modes.
As our observations comparing $x$ = 0 to $x$ = 3 Ca concentrations in Fig.~\ref{fig:Spec_CaDep} suggest, these contrasting effects are mostly balanced ultimately resulting in a subtle softening of each mode following Ca substitution.
On the other hand, the observed transfer of spectral weight between the two modes can be interpreted as an increase in the dipolar moment associated to the $\beta2$ excitation. This effect would be expected following an increase of charge on each sublattice as a result of hole transfer between the chains and ladders as well as from an increase in the relative charge displacement. Since Ca doping leads to an increase in the lattice parameter $c_{\mathrm{ch}}$ but a decrease in $c_{\mathrm{ld}}$, we can indeed expect an increase in the relative charge displacement for a mode where the chains and ladders are displaced in opposite directions.
This picture is also consistent with the temperature dependent observations in Fig.~\ref{fig:Spec_TempDep} (c) for the $x$ = 3 sample. Since charges are more mobile on the ladders than on the chains \cite{Nagata1998} one could expect excitations associated with charge localization on the ladders ($\beta2$) to be suppressed in contrast to chain based modes, as the conductivity on the ladders increases at a higher rate than on the chains when subjected to heating. 

As a final comment, it is interesting to consider the nature of the sliding lattice dynamics in the context of the superconducting state observed at Ca concentrations of approximately $11<x<14$ under pressures on the order of $\sim$3-5 Gpa below $\sim$14 K \cite{Uehara1996,Isobe1998, Radheep2013}. It has recently been commented that low frequency vibrational excitations in other quasi-periodic systems could be significant in relation to observations of strong-coupling superconductivity \cite{Klintberg2012,Tompsett2014,Khasanov2018,Santoro2018,Brown2018,Cheung2018}.
This is because the electron-phonon coupling is essentially proportional to the inverse of phonon frequency ($\lambda\sim\Sigma_q\omega_q^{-1}$) \cite{Brown2018}. Our observed softening of the sliding mode gap suggests that such coupling in SCCO could indeed grow for increasing Ca substitution. While we note that the superconducting channels in SCCO are predicted to occur primarily within the ladders \cite{Dagotto1992,Dagotto1999}, it is interesting to point out that the approximate trend of the energy and intensity of $\beta1$ associated with both slow chain and ladder oscillations approaches zero in the $11<x<14$ superconducting range. This trend, accompanied by a transfer of spectral weight to the fast oscillations associated to $\beta2$, provides a promising indicator that the sliding mode dynamics of SCCO should perhaps be considered within models of the superconducting state.
We also note observations in the literature of CDW softening for increasing Ca substitution in SCCO \cite{Osafune1999,Vuletic2003} which predict a softening to zero energy for Ca concentrations in the range of $x\approx$10. Evidently there is a number of different collective dynamics converging at the phase boundaries of SCCO that indicate collective excitations do play a role in the superconducting thermodynamic landscape.
We anticipate pressure dependent measurements of the sliding mode gap in SCCO to be a high priority in further understanding their overall significance.

\section{Conclusion \label{sec:Concltn}}

In conclusion, we have probed the low energy lattice dynamics of the quasi-periodic spin-ladder cuprate SCCO using synchrotron and laboratory-based THz spectroscopy. Specific focus on the energy gap of quasi-acoustic sliding modes reveals two different excitations we associate with slow in-phase and fast out-of-phase chain-ladder sliding dynamics. An analysis of the dependence on Ca concentration of these modes shows a softening of the sliding mode energy gap and a transfer of spectral weight from the slow oscillations to the fast ones with increasing Ca substitution. The results are consistent with a transfer of intrinsic holes from the chain to ladder sublattices. An approximate trend of the behavior suggests complete suppression of the slower oscillations at the phase boundary of the superconducting phase. The results are discussed in the context of recent interest into the significance of low frequency excitations in aperiodic systems exhibiting superconductivity.
We highlight the importance of future experiments aimed at probing the external pressure dependence of these exotic excitations.


\section {Acknowledgements}

 This work was financially supported in part by Grant No. ANR-13-BS04-0013 and the Swiss State Secretariat of Education and Research (Contract No. JRP122960). E. Constable has benefited from a PRESTIGE (No.2014-1-0020) fellowship and CMIRA funding during this research work.


\begin{thebibliography}{69}%
\makeatletter
\providecommand \@ifxundefined [1]{%
 \@ifx{#1\undefined}
}%
\providecommand \@ifnum [1]{%
 \ifnum #1\expandafter \@firstoftwo
 \else \expandafter \@secondoftwo
 \fi
}%
\providecommand \@ifx [1]{%
 \ifx #1\expandafter \@firstoftwo
 \else \expandafter \@secondoftwo
 \fi
}%
\providecommand \natexlab [1]{#1}%
\providecommand \enquote  [1]{``#1''}%
\providecommand \bibnamefont  [1]{#1}%
\providecommand \bibfnamefont [1]{#1}%
\providecommand \citenamefont [1]{#1}%
\providecommand \href@noop [0]{\@secondoftwo}%
\providecommand \href [0]{\begingroup \@sanitize@url \@href}%
\providecommand \@href[1]{\@@startlink{#1}\@@href}%
\providecommand \@@href[1]{\endgroup#1\@@endlink}%
\providecommand \@sanitize@url [0]{\catcode `\\12\catcode `\$12\catcode
  `\&12\catcode `\#12\catcode `\^12\catcode `\_12\catcode `\%12\relax}%
\providecommand \@@startlink[1]{}%
\providecommand \@@endlink[0]{}%
\providecommand \url  [0]{\begingroup\@sanitize@url \@url }%
\providecommand \@url [1]{\endgroup\@href {#1}{\urlprefix }}%
\providecommand \urlprefix  [0]{URL }%
\providecommand \Eprint [0]{\href }%
\providecommand \doibase [0]{http://dx.doi.org/}%
\providecommand \selectlanguage [0]{\@gobble}%
\providecommand \bibinfo  [0]{\@secondoftwo}%
\providecommand \bibfield  [0]{\@secondoftwo}%
\providecommand \translation [1]{[#1]}%
\providecommand \BibitemOpen [0]{}%
\providecommand \bibitemStop [0]{}%
\providecommand \bibitemNoStop [0]{.\EOS\space}%
\providecommand \EOS [0]{\spacefactor3000\relax}%
\providecommand \BibitemShut  [1]{\csname bibitem#1\endcsname}%
\let\auto@bib@innerbib\@empty
\bibitem [{\citenamefont {Johnston}(2010)}]{Johnston2010}%
  \BibitemOpen
  \bibfield  {author} {\bibinfo {author} {\bibfnamefont {D.~C.}\ \bibnamefont
  {Johnston}},\ }\href@noop {} {\bibfield  {journal} {\bibinfo  {journal} {Adv.
  Phys.}\ }\textbf {\bibinfo {volume} {59}},\ \bibinfo {pages} {803} (\bibinfo
  {year} {2010})}\BibitemShut {NoStop}%
\bibitem [{\citenamefont {Gallais}\ and\ \citenamefont
  {Paul}(2016)}]{Gallais2016}%
  \BibitemOpen
  \bibfield  {author} {\bibinfo {author} {\bibfnamefont {Y.}~\bibnamefont
  {Gallais}}\ and\ \bibinfo {author} {\bibfnamefont {I.}~\bibnamefont {Paul}},\
  }\href {\doibase 10.1016/j.crhy.2015.10.001} {\bibfield  {journal} {\bibinfo
  {journal} {Comptes Rendus Phys.}\ }\textbf {\bibinfo {volume} {17}},\
  \bibinfo {pages} {113} (\bibinfo {year} {2016})}\BibitemShut {NoStop}%
\bibitem [{\citenamefont {Manzeli}\ \emph {et~al.}(2017)\citenamefont
  {Manzeli}, \citenamefont {Ovchinnikov}, \citenamefont {Pasquier},
  \citenamefont {Yazyev},\ and\ \citenamefont {Kis}}]{Manzeli2017}%
  \BibitemOpen
  \bibfield  {author} {\bibinfo {author} {\bibfnamefont {S.}~\bibnamefont
  {Manzeli}}, \bibinfo {author} {\bibfnamefont {D.}~\bibnamefont
  {Ovchinnikov}}, \bibinfo {author} {\bibfnamefont {D.}~\bibnamefont
  {Pasquier}}, \bibinfo {author} {\bibfnamefont {O.~V.}\ \bibnamefont
  {Yazyev}}, \ and\ \bibinfo {author} {\bibfnamefont {A.}~\bibnamefont {Kis}},\
  }\href@noop {} {\bibfield  {journal} {\bibinfo  {journal} {Nat. Rev. Mater.}\
  }\textbf {\bibinfo {volume} {2}},\ \bibinfo {pages} {17033} (\bibinfo {year}
  {2017})}\BibitemShut {NoStop}%
\bibitem [{\citenamefont {Klintberg}\ \emph {et~al.}(2012)\citenamefont
  {Klintberg}, \citenamefont {Goh}, \citenamefont {Alireza}, \citenamefont
  {Saines}, \citenamefont {Tompsett}, \citenamefont {Logg}, \citenamefont
  {Yang}, \citenamefont {Chen}, \citenamefont {Yoshimura},\ and\ \citenamefont
  {Grosche}}]{Klintberg2012}%
  \BibitemOpen
  \bibfield  {author} {\bibinfo {author} {\bibfnamefont {L.~E.}\ \bibnamefont
  {Klintberg}}, \bibinfo {author} {\bibfnamefont {S.~K.}\ \bibnamefont {Goh}},
  \bibinfo {author} {\bibfnamefont {P.~L.}\ \bibnamefont {Alireza}}, \bibinfo
  {author} {\bibfnamefont {P.~J.}\ \bibnamefont {Saines}}, \bibinfo {author}
  {\bibfnamefont {D.~A.}\ \bibnamefont {Tompsett}}, \bibinfo {author}
  {\bibfnamefont {P.~W.}\ \bibnamefont {Logg}}, \bibinfo {author}
  {\bibfnamefont {J.}~\bibnamefont {Yang}}, \bibinfo {author} {\bibfnamefont
  {B.}~\bibnamefont {Chen}}, \bibinfo {author} {\bibfnamefont {K.}~\bibnamefont
  {Yoshimura}}, \ and\ \bibinfo {author} {\bibfnamefont {F.~M.}\ \bibnamefont
  {Grosche}},\ }\href {\doibase 10.1103/physrevlett.109.237008} {\bibfield
  {journal} {\bibinfo  {journal} {Physical Review Letters}\ }\textbf {\bibinfo
  {volume} {109}},\ \bibinfo {pages} {237008} (\bibinfo {year}
  {2012})}\BibitemShut {NoStop}%
\bibitem [{\citenamefont {Tompsett}(2014)}]{Tompsett2014}%
  \BibitemOpen
  \bibfield  {author} {\bibinfo {author} {\bibfnamefont {D.~A.}\ \bibnamefont
  {Tompsett}},\ }\href {\doibase 10.1103/physrevb.89.075117} {\bibfield
  {journal} {\bibinfo  {journal} {Physical Review B}\ }\textbf {\bibinfo
  {volume} {89}},\ \bibinfo {pages} {075117} (\bibinfo {year}
  {2014})}\BibitemShut {NoStop}%
\bibitem [{\citenamefont {Brown}\ \emph {et~al.}(2018)\citenamefont {Brown},
  \citenamefont {Semeniuk}, \citenamefont {Wang}, \citenamefont {Monserrat},
  \citenamefont {Pickard},\ and\ \citenamefont {Grosche}}]{Brown2018}%
  \BibitemOpen
  \bibfield  {author} {\bibinfo {author} {\bibfnamefont {P.}~\bibnamefont
  {Brown}}, \bibinfo {author} {\bibfnamefont {K.}~\bibnamefont {Semeniuk}},
  \bibinfo {author} {\bibfnamefont {D.}~\bibnamefont {Wang}}, \bibinfo {author}
  {\bibfnamefont {B.}~\bibnamefont {Monserrat}}, \bibinfo {author}
  {\bibfnamefont {C.~J.}\ \bibnamefont {Pickard}}, \ and\ \bibinfo {author}
  {\bibfnamefont {F.~M.}\ \bibnamefont {Grosche}},\ }\href {\doibase
  10.1126/sciadv.aao4793} {\bibfield  {journal} {\bibinfo  {journal} {Science
  Advances}\ }\textbf {\bibinfo {volume} {4}},\ \bibinfo {pages} {eaao4793}
  (\bibinfo {year} {2018})}\BibitemShut {NoStop}%
\bibitem [{\citenamefont {Khasanov}\ \emph {et~al.}(2018)\citenamefont
  {Khasanov}, \citenamefont {Luetkens}, \citenamefont {Morenzoni},
  \citenamefont {Simutis}, \citenamefont {Sch{\"o}necker}, \citenamefont
  {{\"O}stlin}, \citenamefont {Chioncel},\ and\ \citenamefont
  {Amato}}]{Khasanov2018}%
  \BibitemOpen
  \bibfield  {author} {\bibinfo {author} {\bibfnamefont {R.}~\bibnamefont
  {Khasanov}}, \bibinfo {author} {\bibfnamefont {H.}~\bibnamefont {Luetkens}},
  \bibinfo {author} {\bibfnamefont {E.}~\bibnamefont {Morenzoni}}, \bibinfo
  {author} {\bibfnamefont {G.}~\bibnamefont {Simutis}}, \bibinfo {author}
  {\bibfnamefont {S.}~\bibnamefont {Sch{\"o}necker}}, \bibinfo {author}
  {\bibfnamefont {A.}~\bibnamefont {{\"O}stlin}}, \bibinfo {author}
  {\bibfnamefont {L.}~\bibnamefont {Chioncel}}, \ and\ \bibinfo {author}
  {\bibfnamefont {A.}~\bibnamefont {Amato}},\ }\href {\doibase
  10.1103/physrevb.98.140504} {\bibfield  {journal} {\bibinfo  {journal}
  {Physical Review B}\ }\textbf {\bibinfo {volume} {98}},\ \bibinfo {pages}
  {140504(R)} (\bibinfo {year} {2018})}\BibitemShut {NoStop}%
\bibitem [{\citenamefont {Santoro}\ \emph {et~al.}(2018)\citenamefont
  {Santoro}, \citenamefont {Colognesi}, \citenamefont {Monserrat},
  \citenamefont {Gregoryanz}, \citenamefont {Ulivi},\ and\ \citenamefont
  {Gorelli}}]{Santoro2018}%
  \BibitemOpen
  \bibfield  {author} {\bibinfo {author} {\bibfnamefont {M.}~\bibnamefont
  {Santoro}}, \bibinfo {author} {\bibfnamefont {D.}~\bibnamefont {Colognesi}},
  \bibinfo {author} {\bibfnamefont {B.}~\bibnamefont {Monserrat}}, \bibinfo
  {author} {\bibfnamefont {E.}~\bibnamefont {Gregoryanz}}, \bibinfo {author}
  {\bibfnamefont {L.}~\bibnamefont {Ulivi}}, \ and\ \bibinfo {author}
  {\bibfnamefont {F.~A.}\ \bibnamefont {Gorelli}},\ }\href {\doibase
  10.1103/physrevb.98.104107} {\bibfield  {journal} {\bibinfo  {journal}
  {Physical Review B}\ }\textbf {\bibinfo {volume} {98}},\ \bibinfo {pages}
  {104107} (\bibinfo {year} {2018})}\BibitemShut {NoStop}%
\bibitem [{\citenamefont {Cheung}\ \emph {et~al.}(2018)\citenamefont {Cheung},
  \citenamefont {Hu}, \citenamefont {Imai}, \citenamefont {Tanioku},
  \citenamefont {Kanagawa}, \citenamefont {Murakawa}, \citenamefont {Moriyama},
  \citenamefont {Zhang}, \citenamefont {Lai}, \citenamefont {Yoshimura},
  \citenamefont {Grosche}, \citenamefont {Kaneko}, \citenamefont {Tsutsui},\
  and\ \citenamefont {Goh}}]{Cheung2018}%
  \BibitemOpen
  \bibfield  {author} {\bibinfo {author} {\bibfnamefont {Y.~W.}\ \bibnamefont
  {Cheung}}, \bibinfo {author} {\bibfnamefont {Y.~J.}\ \bibnamefont {Hu}},
  \bibinfo {author} {\bibfnamefont {M.}~\bibnamefont {Imai}}, \bibinfo {author}
  {\bibfnamefont {Y.}~\bibnamefont {Tanioku}}, \bibinfo {author} {\bibfnamefont
  {H.}~\bibnamefont {Kanagawa}}, \bibinfo {author} {\bibfnamefont
  {J.}~\bibnamefont {Murakawa}}, \bibinfo {author} {\bibfnamefont
  {K.}~\bibnamefont {Moriyama}}, \bibinfo {author} {\bibfnamefont
  {W.}~\bibnamefont {Zhang}}, \bibinfo {author} {\bibfnamefont {K.~T.}\
  \bibnamefont {Lai}}, \bibinfo {author} {\bibfnamefont {K.}~\bibnamefont
  {Yoshimura}}, \bibinfo {author} {\bibfnamefont {F.~M.}\ \bibnamefont
  {Grosche}}, \bibinfo {author} {\bibfnamefont {K.}~\bibnamefont {Kaneko}},
  \bibinfo {author} {\bibfnamefont {S.}~\bibnamefont {Tsutsui}}, \ and\
  \bibinfo {author} {\bibfnamefont {S.~K.}\ \bibnamefont {Goh}},\ }\href
  {\doibase 10.1103/physrevb.98.161103} {\bibfield  {journal} {\bibinfo
  {journal} {Physical Review B}\ }\textbf {\bibinfo {volume} {98}},\ \bibinfo
  {pages} {161103(R)} (\bibinfo {year} {2018})}\BibitemShut {NoStop}%
\bibitem [{\citenamefont {{McCarron III}}\ \emph {et~al.}(1988)\citenamefont
  {{McCarron III}}, \citenamefont {Subramanian}, \citenamefont {Calabrese},\
  and\ \citenamefont {Harlow}}]{Mccarron1988}%
  \BibitemOpen
  \bibfield  {author} {\bibinfo {author} {\bibfnamefont {E.~M.}\ \bibnamefont
  {{McCarron III}}}, \bibinfo {author} {\bibfnamefont {M.~A.}\ \bibnamefont
  {Subramanian}}, \bibinfo {author} {\bibfnamefont {J.~C.}\ \bibnamefont
  {Calabrese}}, \ and\ \bibinfo {author} {\bibfnamefont {R.~L.}\ \bibnamefont
  {Harlow}},\ }\href {\doibase 10.1016/0025-5408(88)90124-9} {\bibfield
  {journal} {\bibinfo  {journal} {Materials Research Bulletin}\ }\textbf
  {\bibinfo {volume} {23}},\ \bibinfo {pages} {1355} (\bibinfo {year}
  {1988})}\BibitemShut {NoStop}%
\bibitem [{\citenamefont {N{\"u}cker}\ \emph {et~al.}(2000)\citenamefont
  {N{\"u}cker}, \citenamefont {Merz}, \citenamefont {Kuntscher}, \citenamefont
  {Gerhold}, \citenamefont {Schuppler}, \citenamefont {Neudert}, \citenamefont
  {Golden}, \citenamefont {Fink}, \citenamefont {Schild}, \citenamefont
  {Stadler},\ and\ \citenamefont {et~al.}}]{Nucker2000}%
  \BibitemOpen
  \bibfield  {author} {\bibinfo {author} {\bibfnamefont {N.}~\bibnamefont
  {N{\"u}cker}}, \bibinfo {author} {\bibfnamefont {M.}~\bibnamefont {Merz}},
  \bibinfo {author} {\bibfnamefont {C.~A.}\ \bibnamefont {Kuntscher}}, \bibinfo
  {author} {\bibfnamefont {S.}~\bibnamefont {Gerhold}}, \bibinfo {author}
  {\bibfnamefont {S.}~\bibnamefont {Schuppler}}, \bibinfo {author}
  {\bibfnamefont {R.}~\bibnamefont {Neudert}}, \bibinfo {author} {\bibfnamefont
  {M.~S.}\ \bibnamefont {Golden}}, \bibinfo {author} {\bibfnamefont
  {J.}~\bibnamefont {Fink}}, \bibinfo {author} {\bibfnamefont {D.}~\bibnamefont
  {Schild}}, \bibinfo {author} {\bibfnamefont {S.}~\bibnamefont {Stadler}}, \
  and\ \bibinfo {author} {\bibnamefont {et~al.}},\ }\href {\doibase
  10.1103/physrevb.62.14384} {\bibfield  {journal} {\bibinfo  {journal}
  {Physical Review B}\ }\textbf {\bibinfo {volume} {62}},\ \bibinfo {pages}
  {14384} (\bibinfo {year} {2000})}\BibitemShut {NoStop}%
\bibitem [{\citenamefont {Akimitsu}\ \emph {et~al.}(1996)\citenamefont
  {Akimitsu}, \citenamefont {Uehara}, \citenamefont {Nagata}, \citenamefont
  {Matsumoto}, \citenamefont {Kitaoka}, \citenamefont {Takahashi},\ and\
  \citenamefont {M\^{o}ri}}]{Akimitsu1996}%
  \BibitemOpen
  \bibfield  {author} {\bibinfo {author} {\bibfnamefont {J.}~\bibnamefont
  {Akimitsu}}, \bibinfo {author} {\bibfnamefont {M.}~\bibnamefont {Uehara}},
  \bibinfo {author} {\bibfnamefont {T.}~\bibnamefont {Nagata}}, \bibinfo
  {author} {\bibfnamefont {S.}~\bibnamefont {Matsumoto}}, \bibinfo {author}
  {\bibfnamefont {Y.}~\bibnamefont {Kitaoka}}, \bibinfo {author} {\bibfnamefont
  {H.}~\bibnamefont {Takahashi}}, \ and\ \bibinfo {author} {\bibfnamefont
  {N.}~\bibnamefont {M\^{o}ri}},\ }\href@noop {} {\bibfield  {journal}
  {\bibinfo  {journal} {Physica C}\ }\textbf {\bibinfo {volume} {263}},\
  \bibinfo {pages} {475} (\bibinfo {year} {1996})}\BibitemShut {NoStop}%
\bibitem [{\citenamefont {Kato}\ \emph {et~al.}(1996)\citenamefont {Kato},
  \citenamefont {Shiota},\ and\ \citenamefont {Koike}}]{Kato1996_2}%
  \BibitemOpen
  \bibfield  {author} {\bibinfo {author} {\bibfnamefont {M.}~\bibnamefont
  {Kato}}, \bibinfo {author} {\bibfnamefont {K.}~\bibnamefont {Shiota}}, \ and\
  \bibinfo {author} {\bibfnamefont {Y.}~\bibnamefont {Koike}},\ }\href@noop {}
  {\bibfield  {journal} {\bibinfo  {journal} {Physica C}\ }\textbf {\bibinfo
  {volume} {258}},\ \bibinfo {pages} {284} (\bibinfo {year}
  {1996})}\BibitemShut {NoStop}%
\bibitem [{\citenamefont {Adachi}\ \emph {et~al.}(1998)\citenamefont {Adachi},
  \citenamefont {Shiota}, \citenamefont {Kato}, \citenamefont {Noji},\ and\
  \citenamefont {Koike}}]{Adachi1998}%
  \BibitemOpen
  \bibfield  {author} {\bibinfo {author} {\bibfnamefont {T.}~\bibnamefont
  {Adachi}}, \bibinfo {author} {\bibfnamefont {K.}~\bibnamefont {Shiota}},
  \bibinfo {author} {\bibfnamefont {M.}~\bibnamefont {Kato}}, \bibinfo {author}
  {\bibfnamefont {T.}~\bibnamefont {Noji}}, \ and\ \bibinfo {author}
  {\bibfnamefont {Y.}~\bibnamefont {Koike}},\ }\href {\doibase
  10.1016/s0038-1098(97)10159-4} {\bibfield  {journal} {\bibinfo  {journal}
  {Solid State Communications}\ }\textbf {\bibinfo {volume} {105}},\ \bibinfo
  {pages} {639} (\bibinfo {year} {1998})}\BibitemShut {NoStop}%
\bibitem [{\citenamefont {Cox}\ \emph {et~al.}(1998)\citenamefont {Cox},
  \citenamefont {Iglesias}, \citenamefont {Hirota}, \citenamefont {Shirane},
  \citenamefont {Matsuda}, \citenamefont {Motoyama}, \citenamefont {Eisaki},\
  and\ \citenamefont {Uchida}}]{Cox1998}%
  \BibitemOpen
  \bibfield  {author} {\bibinfo {author} {\bibfnamefont {D.~E.}\ \bibnamefont
  {Cox}}, \bibinfo {author} {\bibfnamefont {T.}~\bibnamefont {Iglesias}},
  \bibinfo {author} {\bibfnamefont {K.}~\bibnamefont {Hirota}}, \bibinfo
  {author} {\bibfnamefont {G.}~\bibnamefont {Shirane}}, \bibinfo {author}
  {\bibfnamefont {M.}~\bibnamefont {Matsuda}}, \bibinfo {author} {\bibfnamefont
  {N.}~\bibnamefont {Motoyama}}, \bibinfo {author} {\bibfnamefont
  {H.}~\bibnamefont {Eisaki}}, \ and\ \bibinfo {author} {\bibfnamefont
  {S.}~\bibnamefont {Uchida}},\ }\href@noop {} {\bibfield  {journal} {\bibinfo
  {journal} {Physical Review B}\ }\textbf {\bibinfo {volume} {57}},\ \bibinfo
  {pages} {10750} (\bibinfo {year} {1998})}\BibitemShut {NoStop}%
\bibitem [{\citenamefont {Blumberg}\ \emph {et~al.}(2002)\citenamefont
  {Blumberg}, \citenamefont {Littlewood}, \citenamefont {Gozar}, \citenamefont
  {Dennis}, \citenamefont {Motoyama}, \citenamefont {Eisaki},\ and\
  \citenamefont {Uchida}}]{Blumberg2002}%
  \BibitemOpen
  \bibfield  {author} {\bibinfo {author} {\bibfnamefont {G.}~\bibnamefont
  {Blumberg}}, \bibinfo {author} {\bibfnamefont {P.}~\bibnamefont
  {Littlewood}}, \bibinfo {author} {\bibfnamefont {A.}~\bibnamefont {Gozar}},
  \bibinfo {author} {\bibfnamefont {B.~S.}\ \bibnamefont {Dennis}}, \bibinfo
  {author} {\bibfnamefont {N.}~\bibnamefont {Motoyama}}, \bibinfo {author}
  {\bibfnamefont {H.}~\bibnamefont {Eisaki}}, \ and\ \bibinfo {author}
  {\bibfnamefont {S.}~\bibnamefont {Uchida}},\ }\href {\doibase
  10.1126/science.1070481} {\bibfield  {journal} {\bibinfo  {journal}
  {Science}\ }\textbf {\bibinfo {volume} {297}},\ \bibinfo {pages} {584}
  (\bibinfo {year} {2002})}\BibitemShut {NoStop}%
\bibitem [{\citenamefont {Gorshunov}\ \emph {et~al.}(2002)\citenamefont
  {Gorshunov}, \citenamefont {Haas}, \citenamefont {{R\~{o}\~{o}m}},
  \citenamefont {Dressel}, \citenamefont {Vuleti\'{c}}, \citenamefont
  {{Korin-Hamzi\'{c}}}, \citenamefont {Tomi\'{c}}, \citenamefont {Akimitsu},\
  and\ \citenamefont {Nagata}}]{Gorshunov2002}%
  \BibitemOpen
  \bibfield  {author} {\bibinfo {author} {\bibfnamefont {B.}~\bibnamefont
  {Gorshunov}}, \bibinfo {author} {\bibfnamefont {P.}~\bibnamefont {Haas}},
  \bibinfo {author} {\bibfnamefont {T.}~\bibnamefont {{R\~{o}\~{o}m}}},
  \bibinfo {author} {\bibfnamefont {M.}~\bibnamefont {Dressel}}, \bibinfo
  {author} {\bibfnamefont {T.}~\bibnamefont {Vuleti\'{c}}}, \bibinfo {author}
  {\bibfnamefont {B.}~\bibnamefont {{Korin-Hamzi\'{c}}}}, \bibinfo {author}
  {\bibfnamefont {S.}~\bibnamefont {Tomi\'{c}}}, \bibinfo {author}
  {\bibfnamefont {J.}~\bibnamefont {Akimitsu}}, \ and\ \bibinfo {author}
  {\bibfnamefont {T.}~\bibnamefont {Nagata}},\ }\href {\doibase
  10.1103/PhysRevB.66.060508} {\bibfield  {journal} {\bibinfo  {journal}
  {Physical Review B}\ }\textbf {\bibinfo {volume} {66}},\ \bibinfo {pages}
  {060580(R)} (\bibinfo {year} {2002})}\BibitemShut {NoStop}%
\bibitem [{\citenamefont {Choi}\ \emph {et~al.}(2006)\citenamefont {Choi},
  \citenamefont {Grove}, \citenamefont {Lemmens}, \citenamefont {Fischer},
  \citenamefont {G\"{u}ntherodt}, \citenamefont {Ammerahl}, \citenamefont
  {B\"{u}chner}, \citenamefont {Dhalenne}, \citenamefont {Revcolevschi},\ and\
  \citenamefont {Akimitsu}}]{Choi2006}%
  \BibitemOpen
  \bibfield  {author} {\bibinfo {author} {\bibfnamefont {K.-Y.}\ \bibnamefont
  {Choi}}, \bibinfo {author} {\bibfnamefont {M.}~\bibnamefont {Grove}},
  \bibinfo {author} {\bibfnamefont {P.}~\bibnamefont {Lemmens}}, \bibinfo
  {author} {\bibfnamefont {M.}~\bibnamefont {Fischer}}, \bibinfo {author}
  {\bibfnamefont {G.}~\bibnamefont {G\"{u}ntherodt}}, \bibinfo {author}
  {\bibfnamefont {U.}~\bibnamefont {Ammerahl}}, \bibinfo {author}
  {\bibfnamefont {B.}~\bibnamefont {B\"{u}chner}}, \bibinfo {author}
  {\bibfnamefont {G.}~\bibnamefont {Dhalenne}}, \bibinfo {author}
  {\bibfnamefont {A.}~\bibnamefont {Revcolevschi}}, \ and\ \bibinfo {author}
  {\bibfnamefont {J.}~\bibnamefont {Akimitsu}},\ }\href@noop {} {\bibfield
  {journal} {\bibinfo  {journal} {Physical Review B}\ }\textbf {\bibinfo
  {volume} {73}},\ \bibinfo {pages} {104428} (\bibinfo {year}
  {2006})}\BibitemShut {NoStop}%
\bibitem [{\citenamefont {Abbamonte}\ \emph {et~al.}(2004)\citenamefont
  {Abbamonte}, \citenamefont {Blumberg}, \citenamefont {Rusydi}, \citenamefont
  {Gozar}, \citenamefont {Evans}, \citenamefont {Siegrist}, \citenamefont
  {Venema}, \citenamefont {Eisaki}, \citenamefont {Isaacs},\ and\ \citenamefont
  {Sawatzky}}]{Abbamonte2004}%
  \BibitemOpen
  \bibfield  {author} {\bibinfo {author} {\bibfnamefont {P.}~\bibnamefont
  {Abbamonte}}, \bibinfo {author} {\bibfnamefont {G.}~\bibnamefont {Blumberg}},
  \bibinfo {author} {\bibfnamefont {A.}~\bibnamefont {Rusydi}}, \bibinfo
  {author} {\bibfnamefont {A.}~\bibnamefont {Gozar}}, \bibinfo {author}
  {\bibfnamefont {P.~G.}\ \bibnamefont {Evans}}, \bibinfo {author}
  {\bibfnamefont {T.}~\bibnamefont {Siegrist}}, \bibinfo {author}
  {\bibfnamefont {L.}~\bibnamefont {Venema}}, \bibinfo {author} {\bibfnamefont
  {H.}~\bibnamefont {Eisaki}}, \bibinfo {author} {\bibfnamefont {E.~D.}\
  \bibnamefont {Isaacs}}, \ and\ \bibinfo {author} {\bibfnamefont {G.~A.}\
  \bibnamefont {Sawatzky}},\ }\href {\doibase 10.1038/nature02925} {\bibfield
  {journal} {\bibinfo  {journal} {Nature}\ }\textbf {\bibinfo {volume} {431}},\
  \bibinfo {pages} {1078} (\bibinfo {year} {2004})}\BibitemShut {NoStop}%
\bibitem [{\citenamefont {Eccleston}\ \emph {et~al.}(1996)\citenamefont
  {Eccleston}, \citenamefont {Azuma},\ and\ \citenamefont
  {M.}}]{Eccleston1996}%
  \BibitemOpen
  \bibfield  {author} {\bibinfo {author} {\bibfnamefont {R.~S.}\ \bibnamefont
  {Eccleston}}, \bibinfo {author} {\bibfnamefont {M.}~\bibnamefont {Azuma}}, \
  and\ \bibinfo {author} {\bibfnamefont {T.}~\bibnamefont {M.}},\ }\href@noop
  {} {\bibfield  {journal} {\bibinfo  {journal} {Physical review. B, Condensed
  matter}\ }\textbf {\bibinfo {volume} {53}},\ \bibinfo {pages} {R14721}
  (\bibinfo {year} {1996})}\BibitemShut {NoStop}%
\bibitem [{\citenamefont {Matsuda}\ \emph {et~al.}(1996)\citenamefont
  {Matsuda}, \citenamefont {Katsumata}, \citenamefont {Eisaki}, \citenamefont
  {Motoyama}, \citenamefont {Uchida}, \citenamefont {Shapiro},\ and\
  \citenamefont {Shirane}}]{Matsuda1996}%
  \BibitemOpen
  \bibfield  {author} {\bibinfo {author} {\bibfnamefont {M.}~\bibnamefont
  {Matsuda}}, \bibinfo {author} {\bibfnamefont {K.}~\bibnamefont {Katsumata}},
  \bibinfo {author} {\bibfnamefont {H.}~\bibnamefont {Eisaki}}, \bibinfo
  {author} {\bibfnamefont {N.}~\bibnamefont {Motoyama}}, \bibinfo {author}
  {\bibfnamefont {S.}~\bibnamefont {Uchida}}, \bibinfo {author} {\bibfnamefont
  {S.}~\bibnamefont {Shapiro}}, \ and\ \bibinfo {author} {\bibfnamefont
  {G.}~\bibnamefont {Shirane}},\ }\href {\doibase 10.1103/PhysRevB.54.12199}
  {\bibfield  {journal} {\bibinfo  {journal} {Physical Review B}\ }\textbf
  {\bibinfo {volume} {54}},\ \bibinfo {pages} {12199} (\bibinfo {year}
  {1996})}\BibitemShut {NoStop}%
\bibitem [{\citenamefont {Matsuda}\ and\ \citenamefont
  {Katsumata}(1996)}]{Matsuda1996_2}%
  \BibitemOpen
  \bibfield  {author} {\bibinfo {author} {\bibfnamefont {M.}~\bibnamefont
  {Matsuda}}\ and\ \bibinfo {author} {\bibfnamefont {K.}~\bibnamefont
  {Katsumata}},\ }\href@noop {} {\bibfield  {journal} {\bibinfo  {journal}
  {Physical Review B}\ }\textbf {\bibinfo {volume} {53}},\ \bibinfo {pages}
  {12201} (\bibinfo {year} {1996})}\BibitemShut {NoStop}%
\bibitem [{\citenamefont {Eccleston}\ \emph {et~al.}(1998)\citenamefont
  {Eccleston}, \citenamefont {Uehara}, \citenamefont {Akimitsu}, \citenamefont
  {Eisaki}, \citenamefont {Motoyama},\ and\ \citenamefont
  {Uchida}}]{Eccleston1998}%
  \BibitemOpen
  \bibfield  {author} {\bibinfo {author} {\bibfnamefont {R.}~\bibnamefont
  {Eccleston}}, \bibinfo {author} {\bibfnamefont {M.}~\bibnamefont {Uehara}},
  \bibinfo {author} {\bibfnamefont {J.}~\bibnamefont {Akimitsu}}, \bibinfo
  {author} {\bibfnamefont {H.}~\bibnamefont {Eisaki}}, \bibinfo {author}
  {\bibfnamefont {N.}~\bibnamefont {Motoyama}}, \ and\ \bibinfo {author}
  {\bibfnamefont {S.}~\bibnamefont {Uchida}},\ }\href {\doibase
  10.1103/PhysRevLett.81.1702} {\bibfield  {journal} {\bibinfo  {journal}
  {Physical Review Letters}\ }\textbf {\bibinfo {volume} {81}},\ \bibinfo
  {pages} {1702} (\bibinfo {year} {1998})}\BibitemShut {NoStop}%
\bibitem [{\citenamefont {Matsuda}\ \emph {et~al.}(1999)\citenamefont
  {Matsuda}, \citenamefont {Yosihama}, \citenamefont {Kakurai},\ and\
  \citenamefont {Shirane}}]{Matsuda1999}%
  \BibitemOpen
  \bibfield  {author} {\bibinfo {author} {\bibfnamefont {M.}~\bibnamefont
  {Matsuda}}, \bibinfo {author} {\bibfnamefont {T.}~\bibnamefont {Yosihama}},
  \bibinfo {author} {\bibfnamefont {K.}~\bibnamefont {Kakurai}}, \ and\
  \bibinfo {author} {\bibfnamefont {G.}~\bibnamefont {Shirane}},\ }\href@noop
  {} {\bibfield  {journal} {\bibinfo  {journal} {Physical Review B}\ }\textbf
  {\bibinfo {volume} {59}},\ \bibinfo {pages} {1060} (\bibinfo {year}
  {1999})}\BibitemShut {NoStop}%
\bibitem [{\citenamefont {Lorenzo}\ \emph {et~al.}(2010)\citenamefont
  {Lorenzo}, \citenamefont {Regnault}, \citenamefont {Boullier}, \citenamefont
  {Martin}, \citenamefont {Moudden}, \citenamefont {Vanishri}, \citenamefont
  {Marin},\ and\ \citenamefont {Revcolevschi}}]{Lorenzo2010}%
  \BibitemOpen
  \bibfield  {author} {\bibinfo {author} {\bibfnamefont {J.~E.}\ \bibnamefont
  {Lorenzo}}, \bibinfo {author} {\bibfnamefont {L.~P.}\ \bibnamefont
  {Regnault}}, \bibinfo {author} {\bibfnamefont {C.}~\bibnamefont {Boullier}},
  \bibinfo {author} {\bibfnamefont {N.}~\bibnamefont {Martin}}, \bibinfo
  {author} {\bibfnamefont {A.~H.}\ \bibnamefont {Moudden}}, \bibinfo {author}
  {\bibfnamefont {S.}~\bibnamefont {Vanishri}}, \bibinfo {author}
  {\bibfnamefont {C.}~\bibnamefont {Marin}}, \ and\ \bibinfo {author}
  {\bibfnamefont {A.}~\bibnamefont {Revcolevschi}},\ }\href {\doibase
  10.1103/PhysRevLett.105.097202} {\bibfield  {journal} {\bibinfo  {journal}
  {Physical Review Letters}\ }\textbf {\bibinfo {volume} {105}},\ \bibinfo
  {pages} {097202} (\bibinfo {year} {2010})}\BibitemShut {NoStop}%
\bibitem [{\citenamefont {Vuleti\'{c}}\ \emph {et~al.}(2003)\citenamefont
  {Vuleti\'{c}}, \citenamefont {B}, \citenamefont {Tomi\'{c}}, \citenamefont
  {Gorshunov}, \citenamefont {Hass}, \citenamefont {{R\~{o}\~{o}m}},
  \citenamefont {Dressel}, \citenamefont {Akimitsu}, \citenamefont {Sasaki},\
  and\ \citenamefont {Nagata}}]{Vuletic2003}%
  \BibitemOpen
  \bibfield  {author} {\bibinfo {author} {\bibfnamefont {T.}~\bibnamefont
  {Vuleti\'{c}}}, \bibinfo {author} {\bibfnamefont {K.}~\bibnamefont {B}},
  \bibinfo {author} {\bibfnamefont {S.}~\bibnamefont {Tomi\'{c}}}, \bibinfo
  {author} {\bibfnamefont {B.}~\bibnamefont {Gorshunov}}, \bibinfo {author}
  {\bibfnamefont {P.}~\bibnamefont {Hass}}, \bibinfo {author} {\bibfnamefont
  {T.}~\bibnamefont {{R\~{o}\~{o}m}}}, \bibinfo {author} {\bibfnamefont
  {M.}~\bibnamefont {Dressel}}, \bibinfo {author} {\bibfnamefont
  {J.}~\bibnamefont {Akimitsu}}, \bibinfo {author} {\bibfnamefont
  {T.}~\bibnamefont {Sasaki}}, \ and\ \bibinfo {author} {\bibfnamefont
  {T.}~\bibnamefont {Nagata}},\ }\href@noop {} {\bibfield  {journal} {\bibinfo
  {journal} {Physical Review Letters}\ }\textbf {\bibinfo {volume} {90}},\
  \bibinfo {pages} {257002} (\bibinfo {year} {2003})}\BibitemShut {NoStop}%
\bibitem [{\citenamefont {Deng}\ \emph
  {et~al.}(2011{\natexlab{a}})\citenamefont {Deng}, \citenamefont
  {Pomjakushin}, \citenamefont {Pet\v{r}\'{\i}\v{c}ek}, \citenamefont
  {Pomjakushina}, \citenamefont {Kenzelmann},\ and\ \citenamefont
  {Conder}}]{Deng2011_2}%
  \BibitemOpen
  \bibfield  {author} {\bibinfo {author} {\bibfnamefont {G.}~\bibnamefont
  {Deng}}, \bibinfo {author} {\bibfnamefont {V.}~\bibnamefont {Pomjakushin}},
  \bibinfo {author} {\bibfnamefont {V.}~\bibnamefont {Pet\v{r}\'{\i}\v{c}ek}},
  \bibinfo {author} {\bibfnamefont {E.}~\bibnamefont {Pomjakushina}}, \bibinfo
  {author} {\bibfnamefont {M.}~\bibnamefont {Kenzelmann}}, \ and\ \bibinfo
  {author} {\bibfnamefont {K.}~\bibnamefont {Conder}},\ }\href {\doibase
  10.1103/physrevb.84.144111} {\bibfield  {journal} {\bibinfo  {journal}
  {Physical Review B}\ }\textbf {\bibinfo {volume} {84}},\ \bibinfo {pages}
  {144111} (\bibinfo {year} {2011}{\natexlab{a}})}\BibitemShut {NoStop}%
\bibitem [{\citenamefont {Huang}\ \emph {et~al.}(2013)\citenamefont {Huang},
  \citenamefont {Deng}, \citenamefont {Chin}, \citenamefont {Hu}, \citenamefont
  {Cheng}, \citenamefont {Chou}, \citenamefont {Conder}, \citenamefont {Zhou},
  \citenamefont {Pi}, \citenamefont {Goodenough},\ and\ \citenamefont
  {et~al.}}]{Huang2013}%
  \BibitemOpen
  \bibfield  {author} {\bibinfo {author} {\bibfnamefont {M.-J.}\ \bibnamefont
  {Huang}}, \bibinfo {author} {\bibfnamefont {G.}~\bibnamefont {Deng}},
  \bibinfo {author} {\bibfnamefont {Y.~Y.}\ \bibnamefont {Chin}}, \bibinfo
  {author} {\bibfnamefont {Z.}~\bibnamefont {Hu}}, \bibinfo {author}
  {\bibfnamefont {J.-G.}\ \bibnamefont {Cheng}}, \bibinfo {author}
  {\bibfnamefont {F.~C.}\ \bibnamefont {Chou}}, \bibinfo {author}
  {\bibfnamefont {K.}~\bibnamefont {Conder}}, \bibinfo {author} {\bibfnamefont
  {J.-S.}\ \bibnamefont {Zhou}}, \bibinfo {author} {\bibfnamefont {T.-W.}\
  \bibnamefont {Pi}}, \bibinfo {author} {\bibfnamefont {J.~B.}\ \bibnamefont
  {Goodenough}}, \ and\ \bibinfo {author} {\bibnamefont {et~al.}},\ }\href
  {\doibase 10.1103/physrevb.88.014520} {\bibfield  {journal} {\bibinfo
  {journal} {Physical Review B}\ }\textbf {\bibinfo {volume} {88}},\ \bibinfo
  {pages} {014520} (\bibinfo {year} {2013})}\BibitemShut {NoStop}%
\bibitem [{\citenamefont {Deng}\ \emph {et~al.}(2013)\citenamefont {Deng},
  \citenamefont {Tsyrulin}, \citenamefont {Bourges}, \citenamefont {Lamago},
  \citenamefont {Ronnow}, \citenamefont {Kenzelmann}, \citenamefont {Danilkin},
  \citenamefont {Pomjakushina},\ and\ \citenamefont {Conder}}]{Deng2013}%
  \BibitemOpen
  \bibfield  {author} {\bibinfo {author} {\bibfnamefont {G.}~\bibnamefont
  {Deng}}, \bibinfo {author} {\bibfnamefont {N.}~\bibnamefont {Tsyrulin}},
  \bibinfo {author} {\bibfnamefont {P.}~\bibnamefont {Bourges}}, \bibinfo
  {author} {\bibfnamefont {D.}~\bibnamefont {Lamago}}, \bibinfo {author}
  {\bibfnamefont {H.}~\bibnamefont {Ronnow}}, \bibinfo {author} {\bibfnamefont
  {M.}~\bibnamefont {Kenzelmann}}, \bibinfo {author} {\bibfnamefont
  {S.}~\bibnamefont {Danilkin}}, \bibinfo {author} {\bibfnamefont
  {E.}~\bibnamefont {Pomjakushina}}, \ and\ \bibinfo {author} {\bibfnamefont
  {K.}~\bibnamefont {Conder}},\ }\href {\doibase 10.1103/physrevb.88.014504}
  {\bibfield  {journal} {\bibinfo  {journal} {Physical Review B}\ }\textbf
  {\bibinfo {volume} {88}},\ \bibinfo {pages} {014504} (\bibinfo {year}
  {2013})}\BibitemShut {NoStop}%
\bibitem [{\citenamefont {Bag}\ \emph {et~al.}(2018)\citenamefont {Bag},
  \citenamefont {Karmakar}, \citenamefont {Dhar}, \citenamefont {Tripathi},
  \citenamefont {Choudhary},\ and\ \citenamefont {Singh}}]{Bag2018}%
  \BibitemOpen
  \bibfield  {author} {\bibinfo {author} {\bibfnamefont {R.}~\bibnamefont
  {Bag}}, \bibinfo {author} {\bibfnamefont {K.}~\bibnamefont {Karmakar}},
  \bibinfo {author} {\bibfnamefont {S.}~\bibnamefont {Dhar}}, \bibinfo {author}
  {\bibfnamefont {M.}~\bibnamefont {Tripathi}}, \bibinfo {author}
  {\bibfnamefont {R.~J.}\ \bibnamefont {Choudhary}}, \ and\ \bibinfo {author}
  {\bibfnamefont {S.}~\bibnamefont {Singh}},\ }\href {\doibase
  10.1088/1361-648x/aaf01f} {\bibfield  {journal} {\bibinfo  {journal} {Journal
  of Physics: Condensed Matter}\ }\textbf {\bibinfo {volume} {31}},\ \bibinfo
  {pages} {035801} (\bibinfo {year} {2018})}\BibitemShut {NoStop}%
\bibitem [{\citenamefont {Uehara}\ \emph {et~al.}(1996)\citenamefont {Uehara},
  \citenamefont {Nagata}, \citenamefont {Akimitsu}, \citenamefont {Takahashi},
  \citenamefont {M\^{o}ri},\ and\ \citenamefont {Kinoshita}}]{Uehara1996}%
  \BibitemOpen
  \bibfield  {author} {\bibinfo {author} {\bibfnamefont {M.}~\bibnamefont
  {Uehara}}, \bibinfo {author} {\bibfnamefont {T.}~\bibnamefont {Nagata}},
  \bibinfo {author} {\bibfnamefont {J.}~\bibnamefont {Akimitsu}}, \bibinfo
  {author} {\bibfnamefont {H.~.}\ \bibnamefont {Takahashi}}, \bibinfo {author}
  {\bibfnamefont {N.}~\bibnamefont {M\^{o}ri}}, \ and\ \bibinfo {author}
  {\bibfnamefont {K.}~\bibnamefont {Kinoshita}},\ }\href@noop {} {\bibfield
  {journal} {\bibinfo  {journal} {Journal of the Physical Society of Japan}\
  }\textbf {\bibinfo {volume} {65}},\ \bibinfo {pages} {2764} (\bibinfo {year}
  {1996})}\BibitemShut {NoStop}%
\bibitem [{\citenamefont {Isobe}\ \emph {et~al.}(1998)\citenamefont {Isobe},
  \citenamefont {Ohta}, \citenamefont {Onoda}, \citenamefont {Izumi},
  \citenamefont {Nakano}, \citenamefont {Li}, \citenamefont {Matsui},\ and\
  \citenamefont {Takayama-Muromachi}}]{Isobe1998}%
  \BibitemOpen
  \bibfield  {author} {\bibinfo {author} {\bibfnamefont {M.}~\bibnamefont
  {Isobe}}, \bibinfo {author} {\bibfnamefont {T.}~\bibnamefont {Ohta}},
  \bibinfo {author} {\bibfnamefont {M.}~\bibnamefont {Onoda}}, \bibinfo
  {author} {\bibfnamefont {F.}~\bibnamefont {Izumi}}, \bibinfo {author}
  {\bibfnamefont {S.}~\bibnamefont {Nakano}}, \bibinfo {author} {\bibfnamefont
  {J.~Q.}\ \bibnamefont {Li}}, \bibinfo {author} {\bibfnamefont
  {Y.}~\bibnamefont {Matsui}}, \ and\ \bibinfo {author} {\bibfnamefont
  {E.}~\bibnamefont {Takayama-Muromachi}},\ }\href@noop {} {\bibfield
  {journal} {\bibinfo  {journal} {Physical Review B}\ }\textbf {\bibinfo
  {volume} {57}},\ \bibinfo {pages} {613} (\bibinfo {year} {1998})}\BibitemShut
  {NoStop}%
\bibitem [{\citenamefont {Thorsmolle}\ \emph {et~al.}(2012)\citenamefont
  {Thorsmolle}, \citenamefont {Homes}, \citenamefont {Gozar}, \citenamefont
  {Blumberg}, \citenamefont {van Mechelen}, \citenamefont {Kuzmenko},
  \citenamefont {Vanishri}, \citenamefont {Marin},\ and\ \citenamefont
  {Ronnow}}]{Thorsmolle2012}%
  \BibitemOpen
  \bibfield  {author} {\bibinfo {author} {\bibfnamefont {V.~K.}\ \bibnamefont
  {Thorsmolle}}, \bibinfo {author} {\bibfnamefont {C.~C.}\ \bibnamefont
  {Homes}}, \bibinfo {author} {\bibfnamefont {A.}~\bibnamefont {Gozar}},
  \bibinfo {author} {\bibfnamefont {G.}~\bibnamefont {Blumberg}}, \bibinfo
  {author} {\bibfnamefont {J.~L.~M.}\ \bibnamefont {van Mechelen}}, \bibinfo
  {author} {\bibfnamefont {A.~B.}\ \bibnamefont {Kuzmenko}}, \bibinfo {author}
  {\bibfnamefont {S.}~\bibnamefont {Vanishri}}, \bibinfo {author}
  {\bibfnamefont {C.}~\bibnamefont {Marin}}, \ and\ \bibinfo {author}
  {\bibfnamefont {H.~M.}\ \bibnamefont {Ronnow}},\ }\href@noop {} {\bibfield
  {journal} {\bibinfo  {journal} {Physical Review Letters}\ }\textbf {\bibinfo
  {volume} {108}},\ \bibinfo {pages} {217401} (\bibinfo {year}
  {2012})}\BibitemShut {NoStop}%
\bibitem [{\citenamefont {Chen}\ \emph {et~al.}(2016)\citenamefont {Chen},
  \citenamefont {Bansal}, \citenamefont {Sullivan}, \citenamefont {Abernathy},
  \citenamefont {Aczel}, \citenamefont {Zhou}, \citenamefont {Delaire},\ and\
  \citenamefont {Shi}}]{Chen2016}%
  \BibitemOpen
  \bibfield  {author} {\bibinfo {author} {\bibfnamefont {X.}~\bibnamefont
  {Chen}}, \bibinfo {author} {\bibfnamefont {D.}~\bibnamefont {Bansal}},
  \bibinfo {author} {\bibfnamefont {S.}~\bibnamefont {Sullivan}}, \bibinfo
  {author} {\bibfnamefont {D.~L.}\ \bibnamefont {Abernathy}}, \bibinfo {author}
  {\bibfnamefont {A.~A.}\ \bibnamefont {Aczel}}, \bibinfo {author}
  {\bibfnamefont {J.}~\bibnamefont {Zhou}}, \bibinfo {author} {\bibfnamefont
  {O.}~\bibnamefont {Delaire}}, \ and\ \bibinfo {author} {\bibfnamefont
  {L.}~\bibnamefont {Shi}},\ }\href {\doibase 10.1103/physrevb.94.134309}
  {\bibfield  {journal} {\bibinfo  {journal} {Physical Review B}\ }\textbf
  {\bibinfo {volume} {94}},\ \bibinfo {pages} {134309} (\bibinfo {year}
  {2016})}\BibitemShut {NoStop}%
\bibitem [{\citenamefont {Theodorou}\ and\ \citenamefont
  {Rice}(1978)}]{Theodorou1978}%
  \BibitemOpen
  \bibfield  {author} {\bibinfo {author} {\bibfnamefont {G.}~\bibnamefont
  {Theodorou}}\ and\ \bibinfo {author} {\bibfnamefont {T.}~\bibnamefont
  {Rice}},\ }\href@noop {} {\bibfield  {journal} {\bibinfo  {journal} {Physical
  Review B}\ }\textbf {\bibinfo {volume} {18}},\ \bibinfo {pages} {2840}
  (\bibinfo {year} {1978})}\BibitemShut {NoStop}%
\bibitem [{\citenamefont {Theodorou}(1980)}]{Theodorou1980}%
  \BibitemOpen
  \bibfield  {author} {\bibinfo {author} {\bibfnamefont {G.}~\bibnamefont
  {Theodorou}},\ }\href {\doibase 10.1016/0038-1098(80)90860-1} {\bibfield
  {journal} {\bibinfo  {journal} {Solid State Communications}\ }\textbf
  {\bibinfo {volume} {33}},\ \bibinfo {pages} {561} (\bibinfo {year}
  {1980})}\BibitemShut {NoStop}%
\bibitem [{\citenamefont {Ruzicka}\ \emph {et~al.}(1998)\citenamefont
  {Ruzicka}, \citenamefont {Degiorgi}, \citenamefont {Ammerahl}, \citenamefont
  {Dhalenne},\ and\ \citenamefont {Revcolevschi}}]{Ruzicka1998}%
  \BibitemOpen
  \bibfield  {author} {\bibinfo {author} {\bibfnamefont {B.}~\bibnamefont
  {Ruzicka}}, \bibinfo {author} {\bibfnamefont {L.}~\bibnamefont {Degiorgi}},
  \bibinfo {author} {\bibfnamefont {U.}~\bibnamefont {Ammerahl}}, \bibinfo
  {author} {\bibfnamefont {G.}~\bibnamefont {Dhalenne}}, \ and\ \bibinfo
  {author} {\bibfnamefont {A.}~\bibnamefont {Revcolevschi}},\ }\href {\doibase
  10.1007/s100510050552} {\bibfield  {journal} {\bibinfo  {journal} {The
  European Physical Journal B}\ }\textbf {\bibinfo {volume} {6}},\ \bibinfo
  {pages} {301} (\bibinfo {year} {1998})}\BibitemShut {NoStop}%
\bibitem [{\citenamefont {Motoyama}\ \emph {et~al.}(1997)\citenamefont
  {Motoyama}, \citenamefont {Osafune}, \citenamefont {Kakeshita}, \citenamefont
  {Eisaki},\ and\ \citenamefont {Uchida}}]{Motoyama1997}%
  \BibitemOpen
  \bibfield  {author} {\bibinfo {author} {\bibfnamefont {N.}~\bibnamefont
  {Motoyama}}, \bibinfo {author} {\bibfnamefont {T.}~\bibnamefont {Osafune}},
  \bibinfo {author} {\bibfnamefont {T.}~\bibnamefont {Kakeshita}}, \bibinfo
  {author} {\bibfnamefont {H.}~\bibnamefont {Eisaki}}, \ and\ \bibinfo {author}
  {\bibfnamefont {S.}~\bibnamefont {Uchida}},\ }\href {\doibase
  10.1103/PhysRevB.55.R3386} {\bibfield  {journal} {\bibinfo  {journal}
  {Physical Review B}\ }\textbf {\bibinfo {volume} {55}},\ \bibinfo {pages}
  {R3386} (\bibinfo {year} {1997})}\BibitemShut {NoStop}%
\bibitem [{\citenamefont {Deng}\ \emph {et~al.}(2018)\citenamefont {Deng},
  \citenamefont {Yu}, \citenamefont {Mole}, \citenamefont {Pomjakushina},
  \citenamefont {Conder}, \citenamefont {Kenzelmann}, \citenamefont {Yano},
  \citenamefont {Wang}, \citenamefont {Rule}, \citenamefont {Gardner},
  \citenamefont {Luo}, \citenamefont {Li}, \citenamefont {Ulrich},
  \citenamefont {Imperia}, \citenamefont {Ren}, \citenamefont {Cao},\ and\
  \citenamefont {McIntyre}}]{Deng2018}%
  \BibitemOpen
  \bibfield  {author} {\bibinfo {author} {\bibfnamefont {G.}~\bibnamefont
  {Deng}}, \bibinfo {author} {\bibfnamefont {D.}~\bibnamefont {Yu}}, \bibinfo
  {author} {\bibfnamefont {R.}~\bibnamefont {Mole}}, \bibinfo {author}
  {\bibfnamefont {E.}~\bibnamefont {Pomjakushina}}, \bibinfo {author}
  {\bibfnamefont {K.}~\bibnamefont {Conder}}, \bibinfo {author} {\bibfnamefont
  {M.}~\bibnamefont {Kenzelmann}}, \bibinfo {author} {\bibfnamefont {S.-i.}\
  \bibnamefont {Yano}}, \bibinfo {author} {\bibfnamefont {C.-W.}\ \bibnamefont
  {Wang}}, \bibinfo {author} {\bibfnamefont {K.~C.}\ \bibnamefont {Rule}},
  \bibinfo {author} {\bibfnamefont {J.~S.}\ \bibnamefont {Gardner}}, \bibinfo
  {author} {\bibfnamefont {H.}~\bibnamefont {Luo}}, \bibinfo {author}
  {\bibfnamefont {S.}~\bibnamefont {Li}}, \bibinfo {author} {\bibfnamefont
  {C.}~\bibnamefont {Ulrich}}, \bibinfo {author} {\bibfnamefont
  {P.}~\bibnamefont {Imperia}}, \bibinfo {author} {\bibfnamefont
  {W.}~\bibnamefont {Ren}}, \bibinfo {author} {\bibfnamefont {S.}~\bibnamefont
  {Cao}}, \ and\ \bibinfo {author} {\bibfnamefont {G.~J.}\ \bibnamefont
  {McIntyre}},\ }\href {\doibase 10.1103/physrevb.98.184411} {\bibfield
  {journal} {\bibinfo  {journal} {Physical Review B}\ }\textbf {\bibinfo
  {volume} {98}},\ \bibinfo {pages} {184411} (\bibinfo {year}
  {2018})}\BibitemShut {NoStop}%
\bibitem [{\citenamefont {Deng}\ \emph
  {et~al.}(2011{\natexlab{b}})\citenamefont {Deng}, \citenamefont {Radheep},
  \citenamefont {Thiyagarajan}, \citenamefont {Pomjakushina}, \citenamefont
  {Wang}, \citenamefont {Nikseresht}, \citenamefont {Arumugam},\ and\
  \citenamefont {Conder}}]{Deng2011}%
  \BibitemOpen
  \bibfield  {author} {\bibinfo {author} {\bibfnamefont {G.}~\bibnamefont
  {Deng}}, \bibinfo {author} {\bibfnamefont {D.~M.}\ \bibnamefont {Radheep}},
  \bibinfo {author} {\bibfnamefont {R.}~\bibnamefont {Thiyagarajan}}, \bibinfo
  {author} {\bibfnamefont {E.}~\bibnamefont {Pomjakushina}}, \bibinfo {author}
  {\bibfnamefont {S.}~\bibnamefont {Wang}}, \bibinfo {author} {\bibfnamefont
  {N.}~\bibnamefont {Nikseresht}}, \bibinfo {author} {\bibfnamefont
  {S.}~\bibnamefont {Arumugam}}, \ and\ \bibinfo {author} {\bibfnamefont
  {K.}~\bibnamefont {Conder}},\ }\href {\doibase
  10.1016/j.jcrysgro.2011.04.010} {\bibfield  {journal} {\bibinfo  {journal}
  {Journal of Crystal Growth}\ }\textbf {\bibinfo {volume} {327}},\ \bibinfo
  {pages} {182} (\bibinfo {year} {2011}{\natexlab{b}})}\BibitemShut {NoStop}%
\bibitem [{\citenamefont {Barros}\ \emph {et~al.}(2013)\citenamefont {Barros},
  \citenamefont {Evain}, \citenamefont {Manceron}, \citenamefont {Brubach},
  \citenamefont {Tordeux}, \citenamefont {Brunelle}, \citenamefont {Nadolski},
  \citenamefont {Loulergue}, \citenamefont {Couprie}, \citenamefont
  {Bielawski}, \citenamefont {Szwaj},\ and\ \citenamefont {Roy}}]{Barros2013}%
  \BibitemOpen
  \bibfield  {author} {\bibinfo {author} {\bibfnamefont {J.}~\bibnamefont
  {Barros}}, \bibinfo {author} {\bibfnamefont {C.}~\bibnamefont {Evain}},
  \bibinfo {author} {\bibfnamefont {L.}~\bibnamefont {Manceron}}, \bibinfo
  {author} {\bibfnamefont {J.-B.}\ \bibnamefont {Brubach}}, \bibinfo {author}
  {\bibfnamefont {M.-A.}\ \bibnamefont {Tordeux}}, \bibinfo {author}
  {\bibfnamefont {P.}~\bibnamefont {Brunelle}}, \bibinfo {author}
  {\bibfnamefont {L.}~\bibnamefont {Nadolski}}, \bibinfo {author}
  {\bibfnamefont {A.}~\bibnamefont {Loulergue}}, \bibinfo {author}
  {\bibfnamefont {M.-E.}\ \bibnamefont {Couprie}}, \bibinfo {author}
  {\bibfnamefont {S.}~\bibnamefont {Bielawski}}, \bibinfo {author}
  {\bibfnamefont {C.}~\bibnamefont {Szwaj}}, \ and\ \bibinfo {author}
  {\bibfnamefont {P.}~\bibnamefont {Roy}},\ }\href {\doibase 10.1063/1.4793558}
  {\bibfield  {journal} {\bibinfo  {journal} {Review of Scientific
  Instruments}\ }\textbf {\bibinfo {volume} {84}},\ \bibinfo {pages} {033102}
  (\bibinfo {year} {2013})}\BibitemShut {NoStop}%
\bibitem [{\citenamefont {Barros}\ \emph {et~al.}(2015)\citenamefont {Barros},
  \citenamefont {Evain}, \citenamefont {Roussel}, \citenamefont {Manceron},
  \citenamefont {Brubach}, \citenamefont {Tordeux}, \citenamefont {Couprie},
  \citenamefont {Bielawski}, \citenamefont {Szwaj}, \citenamefont {Labat},\
  and\ \citenamefont {et~al.}}]{Barros2015}%
  \BibitemOpen
  \bibfield  {author} {\bibinfo {author} {\bibfnamefont {J.}~\bibnamefont
  {Barros}}, \bibinfo {author} {\bibfnamefont {C.}~\bibnamefont {Evain}},
  \bibinfo {author} {\bibfnamefont {E.}~\bibnamefont {Roussel}}, \bibinfo
  {author} {\bibfnamefont {L.}~\bibnamefont {Manceron}}, \bibinfo {author}
  {\bibfnamefont {J.-B.}\ \bibnamefont {Brubach}}, \bibinfo {author}
  {\bibfnamefont {M.-A.}\ \bibnamefont {Tordeux}}, \bibinfo {author}
  {\bibfnamefont {M.-E.}\ \bibnamefont {Couprie}}, \bibinfo {author}
  {\bibfnamefont {S.}~\bibnamefont {Bielawski}}, \bibinfo {author}
  {\bibfnamefont {C.}~\bibnamefont {Szwaj}}, \bibinfo {author} {\bibfnamefont
  {M.}~\bibnamefont {Labat}}, \ and\ \bibinfo {author} {\bibnamefont
  {et~al.}},\ }\href {\doibase 10.1016/j.jms.2015.03.012} {\bibfield  {journal}
  {\bibinfo  {journal} {Journal of Molecular Spectroscopy}\ }\textbf {\bibinfo
  {volume} {315}},\ \bibinfo {pages} {3} (\bibinfo {year} {2015})}\BibitemShut
  {NoStop}%
\bibitem [{Sup()}]{SupMat}%
  \BibitemOpen
  \href@noop {} {\enquote {\bibinfo {title} {See supplementary materials},}\
  }\BibitemShut {NoStop}%
\bibitem [{\citenamefont {Osafune}\ \emph {et~al.}(1999)\citenamefont
  {Osafune}, \citenamefont {Motoyama}, \citenamefont {Eisaki}, \citenamefont
  {Uchida},\ and\ \citenamefont {Tajima}}]{Osafune1999}%
  \BibitemOpen
  \bibfield  {author} {\bibinfo {author} {\bibfnamefont {T.}~\bibnamefont
  {Osafune}}, \bibinfo {author} {\bibfnamefont {N.}~\bibnamefont {Motoyama}},
  \bibinfo {author} {\bibfnamefont {H.}~\bibnamefont {Eisaki}}, \bibinfo
  {author} {\bibfnamefont {S.}~\bibnamefont {Uchida}}, \ and\ \bibinfo {author}
  {\bibfnamefont {S.}~\bibnamefont {Tajima}},\ }\href {\doibase
  10.1103/PhysRevLett.82.1313} {\bibfield  {journal} {\bibinfo  {journal}
  {Physical Review Letters}\ }\textbf {\bibinfo {volume} {82}},\ \bibinfo
  {pages} {1313} (\bibinfo {year} {1999})}\BibitemShut {NoStop}%
\bibitem [{\citenamefont {Homes}\ \emph {et~al.}(1993)\citenamefont {Homes},
  \citenamefont {Homes}, \citenamefont {Timusk}, \citenamefont {Liang},
  \citenamefont {Bonn},\ and\ \citenamefont {Hardy}}]{Homes1993}%
  \BibitemOpen
  \bibfield  {author} {\bibinfo {author} {\bibfnamefont {C.~C.}\ \bibnamefont
  {Homes}}, \bibinfo {author} {\bibfnamefont {T.~T. C.~C.}\ \bibnamefont
  {Homes}}, \bibinfo {author} {\bibfnamefont {T.}~\bibnamefont {Timusk}},
  \bibinfo {author} {\bibfnamefont {R.}~\bibnamefont {Liang}}, \bibinfo
  {author} {\bibfnamefont {D.~A.}\ \bibnamefont {Bonn}}, \ and\ \bibinfo
  {author} {\bibfnamefont {W.~N.}\ \bibnamefont {Hardy}},\ }\href@noop {}
  {\bibfield  {journal} {\bibinfo  {journal} {Physical review letters}\
  }\textbf {\bibinfo {volume} {71}},\ \bibinfo {pages} {1645} (\bibinfo {year}
  {1993})}\BibitemShut {NoStop}%
\bibitem [{\citenamefont {Gr\"{u}ningera}\ \emph {et~al.}(1998)\citenamefont
  {Gr\"{u}ningera}, \citenamefont {van~der Mare}, \citenamefont {Geserichb},
  \citenamefont {Wolf}, \citenamefont {Erb},\ and\ \citenamefont
  {Kopp}}]{Gruninger1998}%
  \BibitemOpen
  \bibfield  {author} {\bibinfo {author} {\bibfnamefont {M.}~\bibnamefont
  {Gr\"{u}ningera}}, \bibinfo {author} {\bibfnamefont {D.}~\bibnamefont
  {van~der Mare}}, \bibinfo {author} {\bibfnamefont {H.}~\bibnamefont
  {Geserichb}}, \bibinfo {author} {\bibfnamefont {T.}~\bibnamefont {Wolf}},
  \bibinfo {author} {\bibfnamefont {A.}~\bibnamefont {Erb}}, \ and\ \bibinfo
  {author} {\bibfnamefont {T.}~\bibnamefont {Kopp}},\ }\href@noop {} {\bibfield
   {journal} {\bibinfo  {journal} {Physica B: Condensed Matter}\ }\textbf
  {\bibinfo {volume} {244}},\ \bibinfo {pages} {60} (\bibinfo {year}
  {1998})}\BibitemShut {NoStop}%
\bibitem [{\citenamefont {{R\~{o}\~{o}m}}(2004)}]{Room2004}%
  \BibitemOpen
  \bibfield  {author} {\bibinfo {author} {\bibfnamefont {T.}~\bibnamefont
  {{R\~{o}\~{o}m}}},\ }\href@noop {} {\bibfield  {journal} {\bibinfo  {journal}
  {Physical Review B}\ }\textbf {\bibinfo {volume} {70}},\ \bibinfo {pages}
  {144417} (\bibinfo {year} {2004})}\BibitemShut {NoStop}%
\bibitem [{\citenamefont {Ortolani}\ \emph {et~al.}(2006)\citenamefont
  {Ortolani}, \citenamefont {Calvani}, \citenamefont {Lupi}, \citenamefont
  {Schade}, \citenamefont {Perla}, \citenamefont {Fujita},\ and\ \citenamefont
  {Yamada}}]{Ortolani2006}%
  \BibitemOpen
  \bibfield  {author} {\bibinfo {author} {\bibfnamefont {M.}~\bibnamefont
  {Ortolani}}, \bibinfo {author} {\bibfnamefont {P.}~\bibnamefont {Calvani}},
  \bibinfo {author} {\bibfnamefont {S.}~\bibnamefont {Lupi}}, \bibinfo {author}
  {\bibfnamefont {U.}~\bibnamefont {Schade}}, \bibinfo {author} {\bibfnamefont
  {A.}~\bibnamefont {Perla}}, \bibinfo {author} {\bibfnamefont
  {M.}~\bibnamefont {Fujita}}, \ and\ \bibinfo {author} {\bibfnamefont
  {K.}~\bibnamefont {Yamada}},\ }\href@noop {} {\bibfield  {journal} {\bibinfo
  {journal} {Physical Review B}\ }\textbf {\bibinfo {volume} {73}},\ \bibinfo
  {pages} {184508} (\bibinfo {year} {2006})}\BibitemShut {NoStop}%
\bibitem [{\citenamefont {Dumm}\ \emph {et~al.}(2006)\citenamefont {Dumm},
  \citenamefont {Abaker}, \citenamefont {Dressel},\ and\ \citenamefont
  {Montgomery}}]{Dumm2006}%
  \BibitemOpen
  \bibfield  {author} {\bibinfo {author} {\bibfnamefont {M.}~\bibnamefont
  {Dumm}}, \bibinfo {author} {\bibfnamefont {M.}~\bibnamefont {Abaker}},
  \bibinfo {author} {\bibfnamefont {M.}~\bibnamefont {Dressel}}, \ and\
  \bibinfo {author} {\bibfnamefont {L.~K.}\ \bibnamefont {Montgomery}},\ }\href
  {\doibase 10.1007/s10909-006-9101-3} {\bibfield  {journal} {\bibinfo
  {journal} {Journal of Low Temperature Physics}\ }\textbf {\bibinfo {volume}
  {142}},\ \bibinfo {pages} {613} (\bibinfo {year} {2006})}\BibitemShut
  {NoStop}%
\bibitem [{\citenamefont {Pustogow}\ \emph {et~al.}(2016)\citenamefont
  {Pustogow}, \citenamefont {Peterseim}, \citenamefont {Kolatschek},
  \citenamefont {Engel},\ and\ \citenamefont {Dressel}}]{Pustogow2016}%
  \BibitemOpen
  \bibfield  {author} {\bibinfo {author} {\bibfnamefont {A.}~\bibnamefont
  {Pustogow}}, \bibinfo {author} {\bibfnamefont {T.}~\bibnamefont {Peterseim}},
  \bibinfo {author} {\bibfnamefont {S.}~\bibnamefont {Kolatschek}}, \bibinfo
  {author} {\bibfnamefont {L.}~\bibnamefont {Engel}}, \ and\ \bibinfo {author}
  {\bibfnamefont {M.}~\bibnamefont {Dressel}},\ }\href {\doibase
  10.1103/physrevb.94.195125} {\bibfield  {journal} {\bibinfo  {journal}
  {Physical Review B}\ }\textbf {\bibinfo {volume} {94}},\ \bibinfo {pages}
  {195125} (\bibinfo {year} {2016})}\BibitemShut {NoStop}%
\bibitem [{\citenamefont {Popovi\'{c}}\ \emph {et~al.}(2000)\citenamefont
  {Popovi\'{c}}, \citenamefont {Konstantinovi\'{c}}, \citenamefont {Ivanov},
  \citenamefont {Khuong}, \citenamefont {Gaji\'{c}}, \citenamefont {Vietkin},\
  and\ \citenamefont {Moshchalkov}}]{Popovic2000}%
  \BibitemOpen
  \bibfield  {author} {\bibinfo {author} {\bibfnamefont {Z.~V.}\ \bibnamefont
  {Popovi\'{c}}}, \bibinfo {author} {\bibfnamefont {M.~J.}\ \bibnamefont
  {Konstantinovi\'{c}}}, \bibinfo {author} {\bibfnamefont {V.~A.}\ \bibnamefont
  {Ivanov}}, \bibinfo {author} {\bibfnamefont {O.~P.}\ \bibnamefont {Khuong}},
  \bibinfo {author} {\bibfnamefont {R.}~\bibnamefont {Gaji\'{c}}}, \bibinfo
  {author} {\bibfnamefont {A.}~\bibnamefont {Vietkin}}, \ and\ \bibinfo
  {author} {\bibfnamefont {V.~V.}\ \bibnamefont {Moshchalkov}},\ }\href@noop {}
  {\bibfield  {journal} {\bibinfo  {journal} {Physical Review B}\ }\textbf
  {\bibinfo {volume} {62}},\ \bibinfo {pages} {4963} (\bibinfo {year}
  {2000})}\BibitemShut {NoStop}%
\bibitem [{\citenamefont {Sugai}\ and\ \citenamefont
  {Suzuki}(2001)}]{Sugai2001}%
  \BibitemOpen
  \bibfield  {author} {\bibinfo {author} {\bibfnamefont {S.}~\bibnamefont
  {Sugai}}\ and\ \bibinfo {author} {\bibfnamefont {M.}~\bibnamefont {Suzuki}},\
  }\href {\doibase 10.1016/S0022-3697(00)00112-8} {\bibfield  {journal}
  {\bibinfo  {journal} {Journal of Physics and Chemistry of Solids}\ }\textbf
  {\bibinfo {volume} {62}},\ \bibinfo {pages} {119} (\bibinfo {year}
  {2001})}\BibitemShut {NoStop}%
\bibitem [{\citenamefont {Braden}\ \emph {et~al.}(2004)\citenamefont {Braden},
  \citenamefont {Etrillard}, \citenamefont {Gukasov}, \citenamefont
  {Ammerahl},\ and\ \citenamefont {Revcolevschi}}]{Braden2004}%
  \BibitemOpen
  \bibfield  {author} {\bibinfo {author} {\bibfnamefont {M.}~\bibnamefont
  {Braden}}, \bibinfo {author} {\bibfnamefont {J.}~\bibnamefont {Etrillard}},
  \bibinfo {author} {\bibfnamefont {A.}~\bibnamefont {Gukasov}}, \bibinfo
  {author} {\bibfnamefont {U.}~\bibnamefont {Ammerahl}}, \ and\ \bibinfo
  {author} {\bibfnamefont {A.}~\bibnamefont {Revcolevschi}},\ }\href {\doibase
  10.1103/PhysRevB.69.214426} {\bibfield  {journal} {\bibinfo  {journal}
  {Physical Review B}\ }\textbf {\bibinfo {volume} {69}},\ \bibinfo {pages}
  {214426} (\bibinfo {year} {2004})}\BibitemShut {NoStop}%
\bibitem [{\citenamefont {Rusydi}\ \emph {et~al.}(2008)\citenamefont {Rusydi},
  \citenamefont {Abbamonte}, \citenamefont {Eisaki}, \citenamefont {Fujimaki},
  \citenamefont {Smadici}, \citenamefont {Motoyama}, \citenamefont {Uchida},
  \citenamefont {Kim}, \citenamefont {{R\:{u}bhausen}},\ and\ \citenamefont
  {Sawatzky}}]{Rusydi2008}%
  \BibitemOpen
  \bibfield  {author} {\bibinfo {author} {\bibfnamefont {A.}~\bibnamefont
  {Rusydi}}, \bibinfo {author} {\bibfnamefont {P.}~\bibnamefont {Abbamonte}},
  \bibinfo {author} {\bibfnamefont {H.}~\bibnamefont {Eisaki}}, \bibinfo
  {author} {\bibfnamefont {Y.}~\bibnamefont {Fujimaki}}, \bibinfo {author}
  {\bibfnamefont {S.}~\bibnamefont {Smadici}}, \bibinfo {author} {\bibfnamefont
  {N.}~\bibnamefont {Motoyama}}, \bibinfo {author} {\bibfnamefont
  {S.}~\bibnamefont {Uchida}}, \bibinfo {author} {\bibfnamefont {Y.-J.}\
  \bibnamefont {Kim}}, \bibinfo {author} {\bibfnamefont {M.}~\bibnamefont
  {{R\:{u}bhausen}}}, \ and\ \bibinfo {author} {\bibfnamefont {G.~A.}\
  \bibnamefont {Sawatzky}},\ }\href {\doibase 10.1103/PhysRevLett.100.036403}
  {\bibfield  {journal} {\bibinfo  {journal} {Physical Review Letters}\
  }\textbf {\bibinfo {volume} {100}},\ \bibinfo {pages} {036403} (\bibinfo
  {year} {2008})}\BibitemShut {NoStop}%
\bibitem [{\citenamefont {Eldridge}\ \emph {et~al.}(1985)\citenamefont
  {Eldridge}, \citenamefont {Homes}, \citenamefont {Bates},\ and\ \citenamefont
  {Bates}}]{Eldridge1985}%
  \BibitemOpen
  \bibfield  {author} {\bibinfo {author} {\bibfnamefont {J.~E.}\ \bibnamefont
  {Eldridge}}, \bibinfo {author} {\bibfnamefont {C.~C.}\ \bibnamefont {Homes}},
  \bibinfo {author} {\bibfnamefont {F.~E.}\ \bibnamefont {Bates}}, \ and\
  \bibinfo {author} {\bibfnamefont {G.~S.}\ \bibnamefont {Bates}},\ }\href@noop
  {} {\bibfield  {journal} {\bibinfo  {journal} {Physical review. B, Condensed
  matter}\ }\textbf {\bibinfo {volume} {32}},\ \bibinfo {pages} {5156}
  (\bibinfo {year} {1985})}\BibitemShut {NoStop}%
\bibitem [{\citenamefont {Homes}\ and\ \citenamefont
  {Eldridge}(1989)}]{Homes1989}%
  \BibitemOpen
  \bibfield  {author} {\bibinfo {author} {\bibfnamefont {C.~C.}\ \bibnamefont
  {Homes}}\ and\ \bibinfo {author} {\bibfnamefont {J.~E.}\ \bibnamefont
  {Eldridge}},\ }\href@noop {} {\bibfield  {journal} {\bibinfo  {journal}
  {Physical review. B, Condensed matter}\ }\textbf {\bibinfo {volume} {40}},\
  \bibinfo {pages} {6138} (\bibinfo {year} {1989})}\BibitemShut {NoStop}%
\bibitem [{\citenamefont {Damascelli}\ \emph {et~al.}(2000)\citenamefont
  {Damascelli}, \citenamefont {van~der Marel}, \citenamefont {Dhalenne},\ and\
  \citenamefont {Revcolevschi}}]{Damascelli2000}%
  \BibitemOpen
  \bibfield  {author} {\bibinfo {author} {\bibfnamefont {A.}~\bibnamefont
  {Damascelli}}, \bibinfo {author} {\bibfnamefont {D.}~\bibnamefont {van~der
  Marel}}, \bibinfo {author} {\bibfnamefont {G.}~\bibnamefont {Dhalenne}}, \
  and\ \bibinfo {author} {\bibfnamefont {A.}~\bibnamefont {Revcolevschi}},\
  }\href {\doibase 10.1103/physrevb.61.12063} {\bibfield  {journal} {\bibinfo
  {journal} {Physical Review B}\ }\textbf {\bibinfo {volume} {61}},\ \bibinfo
  {pages} {12063} (\bibinfo {year} {2000})}\BibitemShut {NoStop}%
\bibitem [{\citenamefont {Gorshunov}\ \emph {et~al.}(2013)\citenamefont
  {Gorshunov}, \citenamefont {Zhukova}, \citenamefont {Torgashev},
  \citenamefont {Kadyrov}, \citenamefont {Motovilova}, \citenamefont
  {Fischgrabe}, \citenamefont {Moshnyaga}, \citenamefont {Zhang}, \citenamefont
  {Kremer}, \citenamefont {Pracht}, \citenamefont {Zapf},\ and\ \citenamefont
  {Dressel}}]{Gorshunov2013}%
  \BibitemOpen
  \bibfield  {author} {\bibinfo {author} {\bibfnamefont {B.}~\bibnamefont
  {Gorshunov}}, \bibinfo {author} {\bibfnamefont {E.}~\bibnamefont {Zhukova}},
  \bibinfo {author} {\bibfnamefont {V.~I.}\ \bibnamefont {Torgashev}}, \bibinfo
  {author} {\bibfnamefont {L.~S.}\ \bibnamefont {Kadyrov}}, \bibinfo {author}
  {\bibfnamefont {E.~A.}\ \bibnamefont {Motovilova}}, \bibinfo {author}
  {\bibfnamefont {F.}~\bibnamefont {Fischgrabe}}, \bibinfo {author}
  {\bibfnamefont {V.}~\bibnamefont {Moshnyaga}}, \bibinfo {author}
  {\bibfnamefont {T.}~\bibnamefont {Zhang}}, \bibinfo {author} {\bibfnamefont
  {R.}~\bibnamefont {Kremer}}, \bibinfo {author} {\bibfnamefont
  {U.}~\bibnamefont {Pracht}}, \bibinfo {author} {\bibfnamefont
  {S.}~\bibnamefont {Zapf}}, \ and\ \bibinfo {author} {\bibfnamefont
  {M.}~\bibnamefont {Dressel}},\ }\href {\doibase 10.1103/PhysRevB.87.245124}
  {\bibfield  {journal} {\bibinfo  {journal} {Physical Review B}\ }\textbf
  {\bibinfo {volume} {87}},\ \bibinfo {pages} {245124} (\bibinfo {year}
  {2013})}\BibitemShut {NoStop}%
\bibitem [{\citenamefont {H\"{u}vonen}\ \emph {et~al.}(2007)\citenamefont
  {H\"{u}vonen}, \citenamefont {Nagel}, \citenamefont {R\~{o}\~{o}m},
  \citenamefont {Haas}, \citenamefont {Dressel}, \citenamefont {Hwang},
  \citenamefont {Timusk}, \citenamefont {Wang},\ and\ \citenamefont
  {Akimitsu}}]{Huvonen2007}%
  \BibitemOpen
  \bibfield  {author} {\bibinfo {author} {\bibfnamefont {D.}~\bibnamefont
  {H\"{u}vonen}}, \bibinfo {author} {\bibfnamefont {U.}~\bibnamefont {Nagel}},
  \bibinfo {author} {\bibfnamefont {T.}~\bibnamefont {R\~{o}\~{o}m}}, \bibinfo
  {author} {\bibfnamefont {P.}~\bibnamefont {Haas}}, \bibinfo {author}
  {\bibfnamefont {M.}~\bibnamefont {Dressel}}, \bibinfo {author} {\bibfnamefont
  {J.}~\bibnamefont {Hwang}}, \bibinfo {author} {\bibfnamefont
  {T.}~\bibnamefont {Timusk}}, \bibinfo {author} {\bibfnamefont {Y.-J.}\
  \bibnamefont {Wang}}, \ and\ \bibinfo {author} {\bibfnamefont
  {J.}~\bibnamefont {Akimitsu}},\ }\href@noop {} {\bibfield  {journal}
  {\bibinfo  {journal} {Physical Review B}\ }\textbf {\bibinfo {volume} {76}},\
  \bibinfo {pages} {134418} (\bibinfo {year} {2007})}\BibitemShut {NoStop}%
\bibitem [{\citenamefont {Constable}\ \emph {et~al.}(2014)\citenamefont
  {Constable}, \citenamefont {Deng}, \citenamefont {Horvat},\ and\
  \citenamefont {Lewis}}]{Constable2014}%
  \BibitemOpen
  \bibfield  {author} {\bibinfo {author} {\bibfnamefont {E.}~\bibnamefont
  {Constable}}, \bibinfo {author} {\bibfnamefont {G.}~\bibnamefont {Deng}},
  \bibinfo {author} {\bibfnamefont {J.}~\bibnamefont {Horvat}}, \ and\ \bibinfo
  {author} {\bibfnamefont {R.~A.}\ \bibnamefont {Lewis}},\ }\href {\doibase
  10.1109/irmmw-thz.2014.6956359} {\bibfield  {journal} {\bibinfo  {journal}
  {2014 39th International Conference on Infrared, Millimeter, and Terahertz
  waves (IRMMW-THz)}\ } (\bibinfo {year} {2014}),\
  10.1109/irmmw-thz.2014.6956359}\BibitemShut {NoStop}%
\bibitem [{\citenamefont {Pimenov}\ \emph {et~al.}(2002)\citenamefont
  {Pimenov}, \citenamefont {Pronin}, \citenamefont {Loidl}, \citenamefont
  {Tsukada},\ and\ \citenamefont {Naito}}]{Pimenov2002}%
  \BibitemOpen
  \bibfield  {author} {\bibinfo {author} {\bibfnamefont {A.}~\bibnamefont
  {Pimenov}}, \bibinfo {author} {\bibfnamefont {A.}~\bibnamefont {Pronin}},
  \bibinfo {author} {\bibfnamefont {A.}~\bibnamefont {Loidl}}, \bibinfo
  {author} {\bibfnamefont {A.}~\bibnamefont {Tsukada}}, \ and\ \bibinfo
  {author} {\bibfnamefont {M.}~\bibnamefont {Naito}},\ }\href@noop {} {\
  \textbf {\bibinfo {volume} {66}},\ \bibinfo {pages} {212508} (\bibinfo {year}
  {2002})}\BibitemShut {NoStop}%
\bibitem [{\citenamefont {Osafune}\ \emph {et~al.}(1997)\citenamefont
  {Osafune}, \citenamefont {Motoyama}, \citenamefont {Eisaki},\ and\
  \citenamefont {Uchida}}]{Osafune1997}%
  \BibitemOpen
  \bibfield  {author} {\bibinfo {author} {\bibfnamefont {T.}~\bibnamefont
  {Osafune}}, \bibinfo {author} {\bibfnamefont {N.}~\bibnamefont {Motoyama}},
  \bibinfo {author} {\bibfnamefont {H.}~\bibnamefont {Eisaki}}, \ and\ \bibinfo
  {author} {\bibfnamefont {S.}~\bibnamefont {Uchida}},\ }\href {\doibase
  10.1103/PhysRevLett.78.1980} {\bibfield  {journal} {\bibinfo  {journal}
  {Physical Review Letters}\ }\textbf {\bibinfo {volume} {78}},\ \bibinfo
  {pages} {1980} (\bibinfo {year} {1997})}\BibitemShut {NoStop}%
\bibitem [{\citenamefont {Schmidt}\ \emph {et~al.}(2003)\citenamefont
  {Schmidt}, \citenamefont {Knetter}, \citenamefont {Gr{\"u}ninger},\ and\
  \citenamefont {Uhrig}}]{Schmidt2003}%
  \BibitemOpen
  \bibfield  {author} {\bibinfo {author} {\bibfnamefont {K.~P.}\ \bibnamefont
  {Schmidt}}, \bibinfo {author} {\bibfnamefont {C.}~\bibnamefont {Knetter}},
  \bibinfo {author} {\bibfnamefont {M.}~\bibnamefont {Gr{\"u}ninger}}, \ and\
  \bibinfo {author} {\bibfnamefont {G.~S.}\ \bibnamefont {Uhrig}},\ }\href
  {\doibase 10.1103/physrevlett.90.167201} {\bibfield  {journal} {\bibinfo
  {journal} {Physical Review Letters}\ }\textbf {\bibinfo {volume} {90}},\
  \bibinfo {pages} {167201} (\bibinfo {year} {2003})}\BibitemShut {NoStop}%
\bibitem [{\citenamefont {Axe}\ and\ \citenamefont {Bak}(1982)}]{Axe1982}%
  \BibitemOpen
  \bibfield  {author} {\bibinfo {author} {\bibfnamefont {J.~D.}\ \bibnamefont
  {Axe}}\ and\ \bibinfo {author} {\bibfnamefont {P.}~\bibnamefont {Bak}},\
  }\href@noop {} {\bibfield  {journal} {\bibinfo  {journal} {Phys. Rev. B}\
  }\textbf {\bibinfo {volume} {26}},\ \bibinfo {pages} {4963} (\bibinfo {year}
  {1982})}\BibitemShut {NoStop}%
\bibitem [{\citenamefont {Osada}\ \emph {et~al.}(2000)\citenamefont {Osada},
  \citenamefont {Kakihana}, \citenamefont {Nagai}, \citenamefont {Noji},
  \citenamefont {Adachi}, \citenamefont {Koike}, \citenamefont
  {B\"{a}ckstr\"{o}m}, \citenamefont {K\"{a}ll},\ and\ \citenamefont
  {B\"{o}rjesson}}]{Osada2000}%
  \BibitemOpen
  \bibfield  {author} {\bibinfo {author} {\bibfnamefont {M.}~\bibnamefont
  {Osada}}, \bibinfo {author} {\bibfnamefont {M.}~\bibnamefont {Kakihana}},
  \bibinfo {author} {\bibfnamefont {I.}~\bibnamefont {Nagai}}, \bibinfo
  {author} {\bibfnamefont {T.}~\bibnamefont {Noji}}, \bibinfo {author}
  {\bibfnamefont {T.}~\bibnamefont {Adachi}}, \bibinfo {author} {\bibfnamefont
  {Y.}~\bibnamefont {Koike}}, \bibinfo {author} {\bibfnamefont
  {J.}~\bibnamefont {B\"{a}ckstr\"{o}m}}, \bibinfo {author} {\bibfnamefont
  {M.}~\bibnamefont {K\"{a}ll}}, \ and\ \bibinfo {author} {\bibfnamefont
  {L.}~\bibnamefont {B\"{o}rjesson}},\ }\href@noop {} {\bibfield  {journal}
  {\bibinfo  {journal} {Physica C}\ }\textbf {\bibinfo {volume} {338}},\
  \bibinfo {pages} {161} (\bibinfo {year} {2000})}\BibitemShut {NoStop}%
\bibitem [{\citenamefont {Nagata}\ \emph {et~al.}(1998)\citenamefont {Nagata},
  \citenamefont {Uehara}, \citenamefont {Goto}, \citenamefont {Akimitsu},
  \citenamefont {Motoyama}, \citenamefont {Eisaki}, \citenamefont {Uchida},
  \citenamefont {Takahashi}, \citenamefont {Nakanishi},\ and\ \citenamefont
  {M\^{o}ri}}]{Nagata1998}%
  \BibitemOpen
  \bibfield  {author} {\bibinfo {author} {\bibfnamefont {T.}~\bibnamefont
  {Nagata}}, \bibinfo {author} {\bibfnamefont {M.}~\bibnamefont {Uehara}},
  \bibinfo {author} {\bibfnamefont {J.}~\bibnamefont {Goto}}, \bibinfo {author}
  {\bibfnamefont {J.}~\bibnamefont {Akimitsu}}, \bibinfo {author}
  {\bibfnamefont {N.}~\bibnamefont {Motoyama}}, \bibinfo {author}
  {\bibfnamefont {H.}~\bibnamefont {Eisaki}}, \bibinfo {author} {\bibfnamefont
  {S.}~\bibnamefont {Uchida}}, \bibinfo {author} {\bibfnamefont
  {H.}~\bibnamefont {Takahashi}}, \bibinfo {author} {\bibfnamefont
  {T.}~\bibnamefont {Nakanishi}}, \ and\ \bibinfo {author} {\bibfnamefont
  {N.}~\bibnamefont {M\^{o}ri}},\ }\href {\doibase 10.1103/PhysRevLett.81.1090}
  {\bibfield  {journal} {\bibinfo  {journal} {Physical Review Letters}\
  }\textbf {\bibinfo {volume} {81}},\ \bibinfo {pages} {1090} (\bibinfo {year}
  {1998})}\BibitemShut {NoStop}%
\bibitem [{\citenamefont {Radheep}\ \emph {et~al.}(2013)\citenamefont
  {Radheep}, \citenamefont {Thiyagarjan}, \citenamefont {Esakkimuthu},
  \citenamefont {Deng}, \citenamefont {Pomjakushina}, \citenamefont {Prajapat},
  \citenamefont {Ravikumar}, \citenamefont {Conder}, \citenamefont {Baskaran},\
  and\ \citenamefont {Arumugam}}]{Radheep2013}%
  \BibitemOpen
  \bibfield  {author} {\bibinfo {author} {\bibfnamefont {D.~M.}\ \bibnamefont
  {Radheep}}, \bibinfo {author} {\bibfnamefont {R.}~\bibnamefont
  {Thiyagarjan}}, \bibinfo {author} {\bibfnamefont {S.}~\bibnamefont
  {Esakkimuthu}}, \bibinfo {author} {\bibfnamefont {G.}~\bibnamefont {Deng}},
  \bibinfo {author} {\bibfnamefont {E.}~\bibnamefont {Pomjakushina}}, \bibinfo
  {author} {\bibfnamefont {C.~L.}\ \bibnamefont {Prajapat}}, \bibinfo {author}
  {\bibfnamefont {G.}~\bibnamefont {Ravikumar}}, \bibinfo {author}
  {\bibfnamefont {K.}~\bibnamefont {Conder}}, \bibinfo {author} {\bibfnamefont
  {G.}~\bibnamefont {Baskaran}}, \ and\ \bibinfo {author} {\bibfnamefont
  {S.}~\bibnamefont {Arumugam}},\ }\href@noop {} {\ ,\ \bibinfo {pages}
  {arXiv:1303.0921} (\bibinfo {year} {2013})}\BibitemShut {NoStop}%
\bibitem [{\citenamefont {Dagotto}\ \emph {et~al.}(1992)\citenamefont
  {Dagotto}, \citenamefont {Riera},\ and\ \citenamefont
  {Scalapino}}]{Dagotto1992}%
  \BibitemOpen
  \bibfield  {author} {\bibinfo {author} {\bibfnamefont {E.}~\bibnamefont
  {Dagotto}}, \bibinfo {author} {\bibfnamefont {J.}~\bibnamefont {Riera}}, \
  and\ \bibinfo {author} {\bibfnamefont {D.}~\bibnamefont {Scalapino}},\
  }\href@noop {} {\bibfield  {journal} {\bibinfo  {journal} {Physical Review
  B}\ }\textbf {\bibinfo {volume} {45}},\ \bibinfo {pages} {5744} (\bibinfo
  {year} {1992})}\BibitemShut {NoStop}%
\bibitem [{\citenamefont {Dagotto}(1999)}]{Dagotto1999}%
  \BibitemOpen
  \bibfield  {author} {\bibinfo {author} {\bibfnamefont {E.}~\bibnamefont
  {Dagotto}},\ }\href@noop {} {\bibfield  {journal} {\bibinfo  {journal}
  {Reports on Progress in Physics}\ }\textbf {\bibinfo {volume} {62}},\
  \bibinfo {pages} {1525} (\bibinfo {year} {1999})}\BibitemShut {NoStop}%
\end{thebibliography}
\end{document}


\title{Supplementary material: Effects of Ca substitution on quasi-acoustic sliding modes in Sr$_{14-\mathrm{x}}$Ca$_{\mathrm{x}}$Cu$_{24}$O$_{41}$}

\author{E. Constable}
\email[]{evan.constable@tuwien.ac.at}
\affiliation{Institute of Solid State Physics, Vienna University of Technology, 1040 Vienna, Austria}
\affiliation{Institut N\'eel, CNRS and Universit\'{e} Grenoble Alpes, 38042 Grenoble, France}
\affiliation{Institute for Superconducting and Electronic Materials, University of Wollongong, Wollongong, NSW 2522, Australia}

\author {A. D. Squires}
\author{J. Horvat}
\author {R. A. Lewis}
\affiliation{Institute for Superconducting and Electronic Materials, University of Wollongong, Wollongong, NSW 2522, Australia}
\author{D. Appadoo}
\author{R. Plathe}
\affiliation{Australian Synchrotron, Australian Nuclear Science and Technology Organisation, 800 Blackburn Rd Clayton, VIC 3168, Australia}
\author {P. Roy}
\author {J.-B. Brubach}
\affiliation{Synchrotron SOLEIL, L'Orme des Merisiers, Saint-Aubin, BP 48, F-91192 Gif-sur-Yvette Cedex, France}
\author {S. deBrion}
\affiliation{Institut N\'eel, CNRS and Universit\'{e} Grenoble Alpes, 38042 Grenoble, France}
\author{A. Pimenov}\affiliation{Institute of Solid State Physics, Vienna University of Technology, 1040 Vienna, Austria}
\author {G. Deng}
\email[]{guochu.deng@ansto.gov.au}
\affiliation{Australian Centre for Neutron Scattering, Australian Nuclear Science and Technology Organisation, New Illawarra Rd, Lucas Heights, Sydney, NSW 2234, Australia}


\maketitle
\onecolumngrid

\section{Temperature dependence of fitted peak parameters between 1-3 THz}


 \begin{figure}[H]
 \centering
 \includegraphics{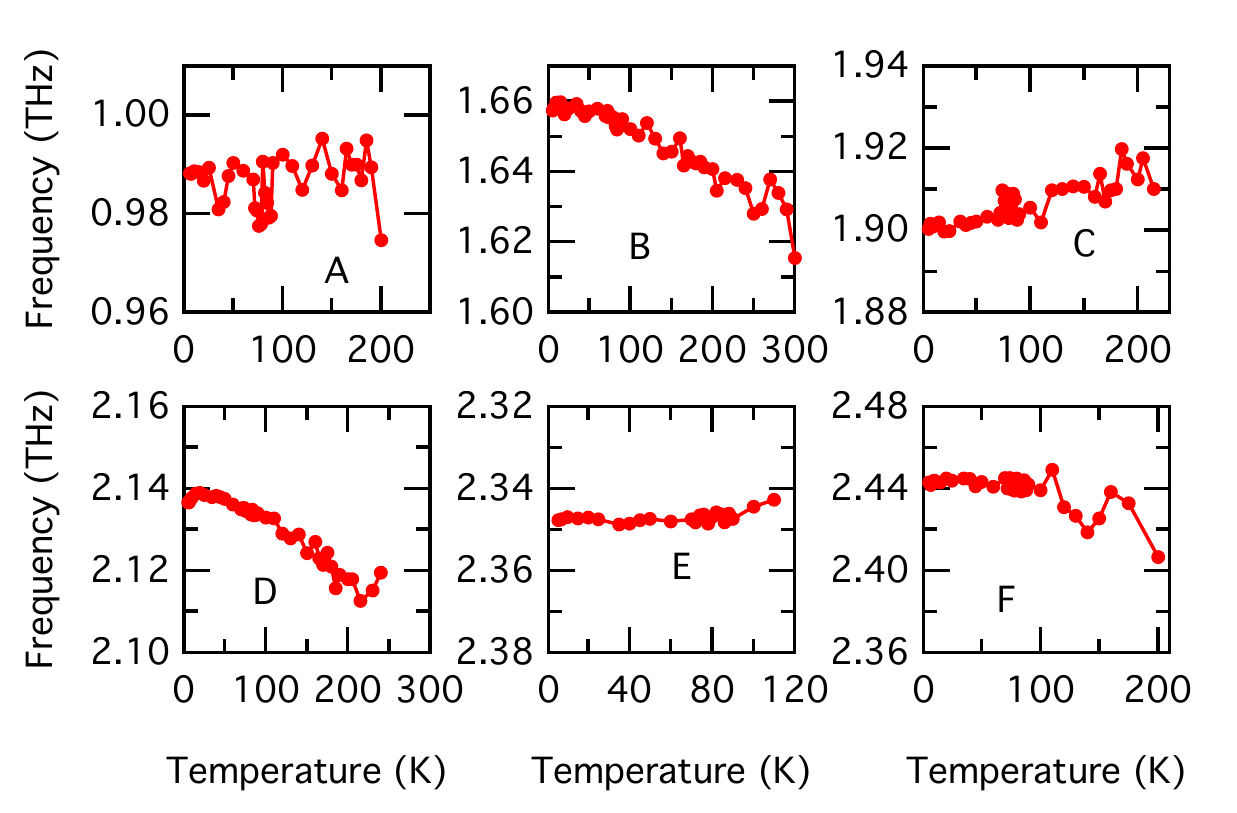}
 \caption{Frequency position for peaks in the $E||a$ configuration from Fig. 2 of the main text}
 \label{fig:Ep2a_posALL}
 \end{figure}
 \begin{figure}[H]
  \centering
  \includegraphics{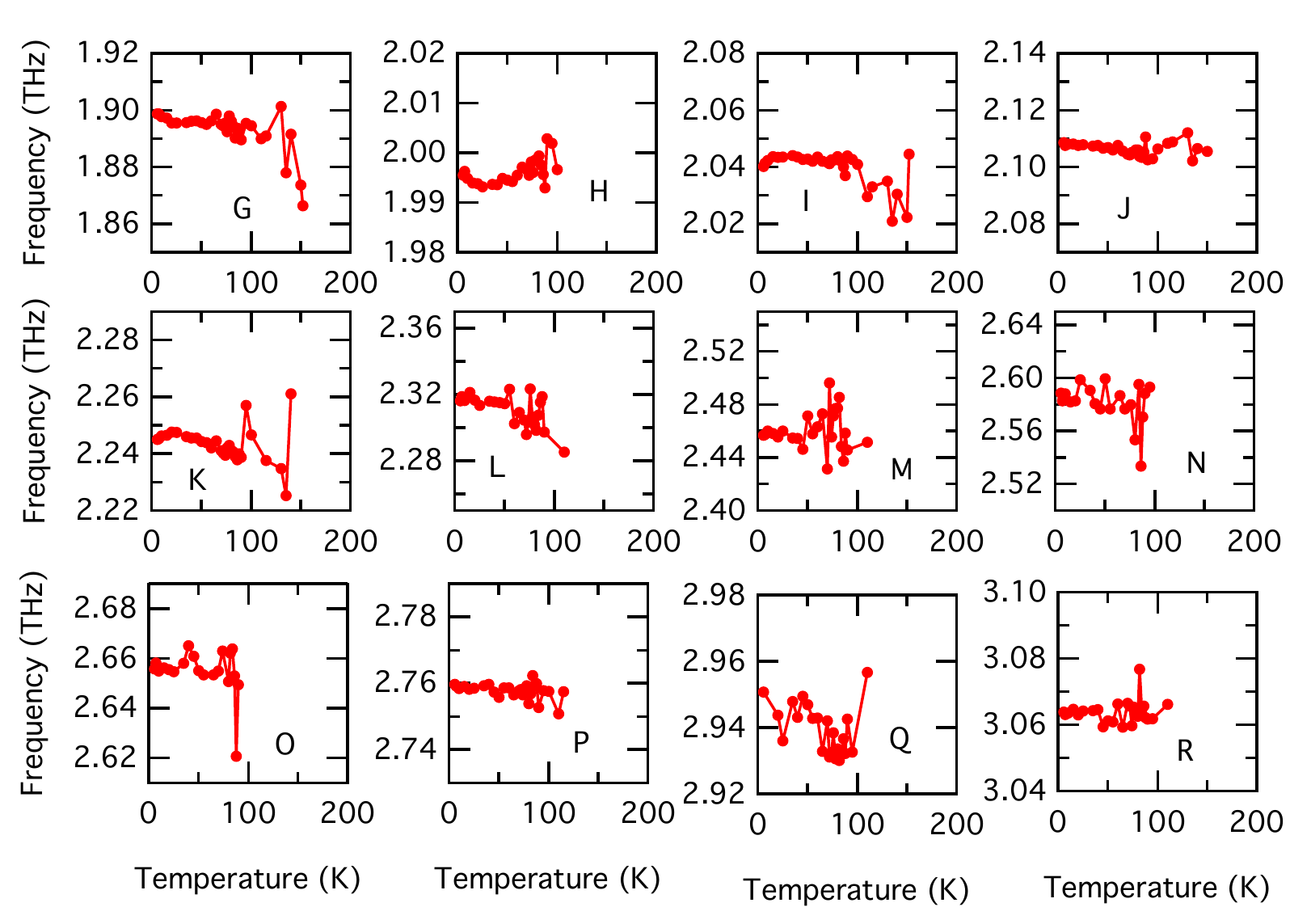}
  \caption{Frequency position for peaks in the $E||c$ configuration from Fig. 2 of the main text}
  \label{fig:Ep2c_posALL}
  \end{figure}
   \begin{figure}[H]
    \centering
    \includegraphics{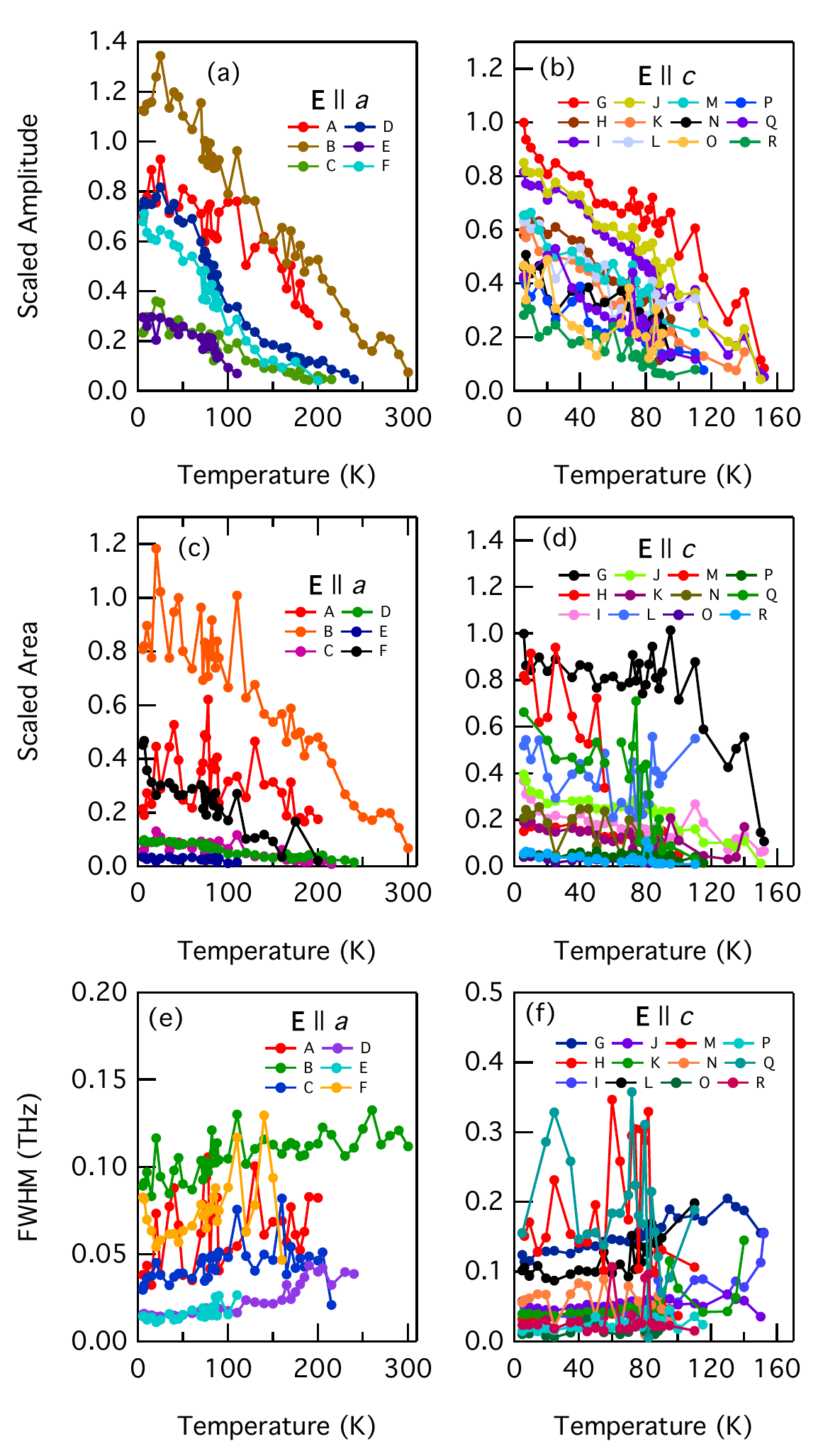}
    \caption{Fitted peak amplitude (a-b), area (c-d) and full width at half maximum (FWHM) (e-f) for peaks in Fig. 2 of the main text. The amplitude and areas are scaled to the base temperature (5.6 K) values of peaks B and G for the $E||a$ and $E||c$ configurations respectively.}
    \label{fig:Peak_params}
    \end{figure}

\section{Coherent synchrotron quantitative data analysis technique}

\begin{figure}
\centering
\includegraphics{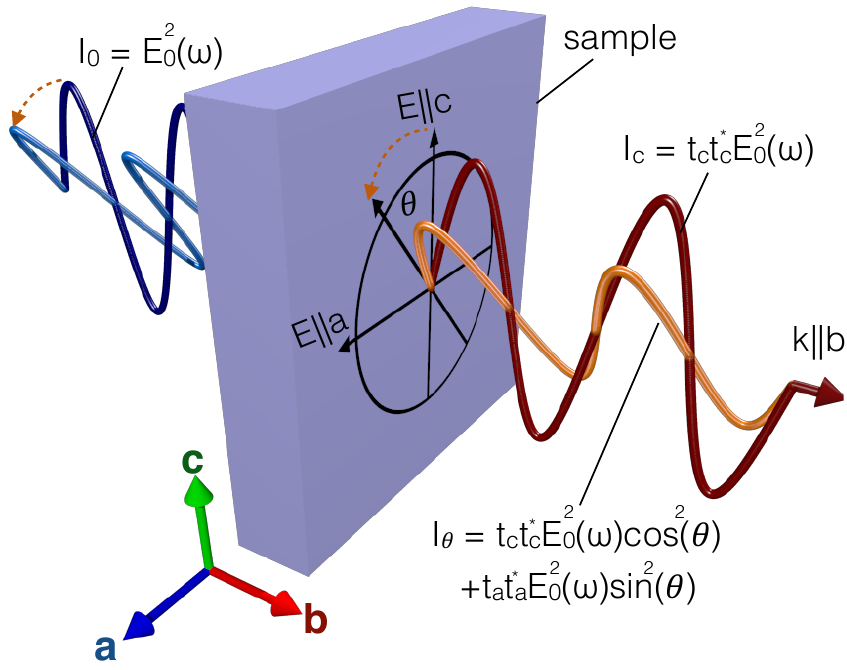}
\caption{Graphical depiction of measurement technique to extract quantitative sample response from rotated spectra in an anisotropic material.}
\label{fig:experiment}
\end{figure}

The measurements using the coherent synchrotron radiation (CSR) source were limited in the ability to obtain a vacuum reference to extract a quantitative sample transmission response. Therefore we have taken advantage of the highly anisotropic optical response of the \SxCO system to extract an approximate absorption response using a rotationally dependent transmission measurement method. An overview of the method is depicted in Fig~\ref{fig:experiment}.
To probe the sample absorption along the $c$ direction radiation polarized parallel to $c$ is directed onto the sample. 
The incident radiation is given by $I_0 = E^2_0(\omega)$. The intensity of the transmitted radiation parallel to $c$ is then $I_c = t_ct_c^*E^2_0(\omega)$ where $t_c$ is the complex transmittance and $t_c^*$ is its complex conjugate.
By rotating the polarization of the radiation relative to the sample at an angle $\theta$, the transmitted intensity now samples a projection on to both the $a$ and $c$ transmittance properties $I_{\theta} = t_ct_c^*E^2_0(\omega)\cos^2(\theta)+t_at_a^*E^2_0(\omega)\sin^2(\theta)$.
Assuming a fixed response for the transmittance along $a$ as indicated by our own studies and previous works, the transmittance properties for the $c$ direction can be extracted independently using the following analysis procedure:

After canceling the common terms of the indecent electric field, a ratio between the transmitted intensity for $E||c$ ($I_c$ ) and that at some angle $\theta$ away from $E||c$ ($I_{\theta}$) is given by:
\begin{equation}
\frac{I_c}{I_{\theta}} = \frac{t_ct_c^*}{t_ct_c^*\cos^2(\theta)+t_at_a^*\sin^2(\theta)},
\end{equation}
such that,
\begin{equation}
\frac{I_{\theta}}{I_c} - \cos^2(\theta) = \frac{t_at_a^*}{t_ct_c^*}\sin^2(\theta).
\end{equation}
Then the transmission coefficient in the $c$ direction is given by,
 \begin{equation}
 t_ct_c^* = t_at_a^*\sin^2(\theta)\left[\frac{I_{\theta}}{I_c} - \cos^2(\theta)\right]^{-1},
 \end{equation}
  and the attenuation coefficient defined by the Beer-Lambert law is
  \begin{equation} 
  \label{Eq:abs}
  \alpha = -\frac{\ln(t_ct_c^*)}{d},
  \end{equation}   
  where $d$ is the sample thickness.

\begin{figure}
\centering
\includegraphics{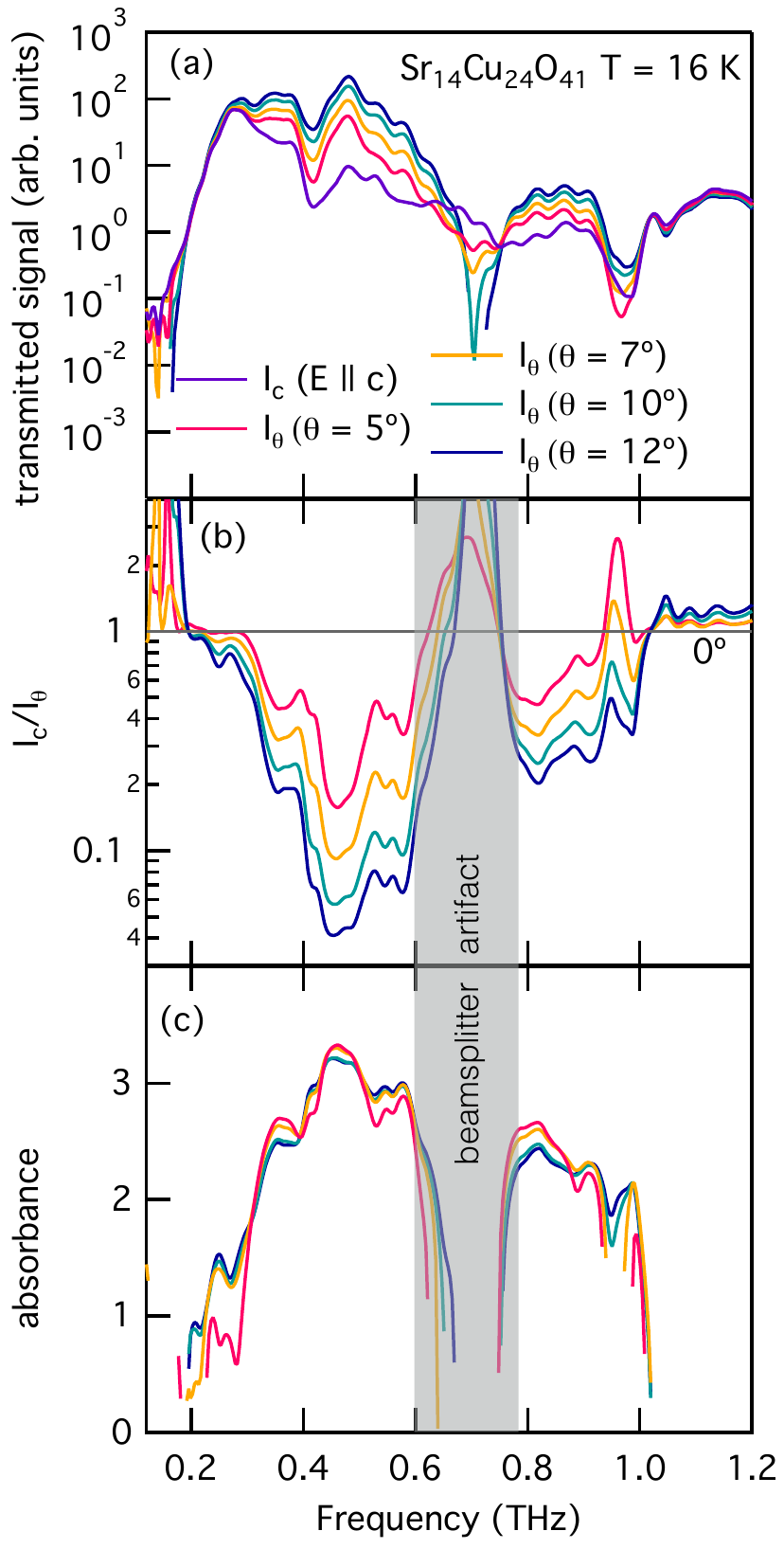}
\caption{(a) Raw spectra of \SCO using CSR source at increasing angles rotated away from a $E||c$ configuration. (b) Ratio of raw spectra at different angles with a measurement for $E||c$. (d) Sample absorption ($\alpha d$) calculated using equation \ref{Eq:abs}. Agreement between the different angles highlights the validity of the analysis method.}
\label{fig:specsexp}
\end{figure}

The independent steps of this analysis procedure are shown in Fig.~\ref{fig:specsexp}. We find that this method is resilient to small errors in the assumed value of $T_a$.
Raw Spectra are shown as the angle of polarization is rotated towards the $E||c$ configuration Fig.~\ref{fig:specsexp} (a). 
Artifacts due to the beamsplitter minimum and the radiation profile are located at 0.70 and 0.96 THz and must be omitted from the final data extraction. An excitation band emerges between 0.2-1 THz when the angle between the relative spectra is increased.
A broad absorption is observed as the sample is aligned to the polarization configuration $E||c$ which is highlight in ratio between the two spectra shown in Fig.~\ref{fig:specsexp} (b). 
One of the benefits of using the CSR low-$\alpha$ mode radiation is the high flux at sub-THz frequencies which are not completely absorbed despite the strong excitations in this energy range. 
The quantitative absorption ($\alpha d$) calculated using equations (3) and (4) is shown in Fig.~\ref{fig:specsexp} (c). 
Here we have used a fixed transmission value for $T_a$ of 0.35 extrapolated from the measurements at higher frequency.
We assume an uncertainty of $\pm$0.1 which translates to a small frequency independent offset.
The agreement between the spectra obtained at different angles demonstrates the validity of this method
Weak fringes appear on top of the excitation band likely due to the a phase mismatch between the plane parallel interference fringes for $E||c$ and $E||a$

\section{Model diagram of proposed slow and fast sliding mode excitations}

\begin{figure}[H]
\centering
\includegraphics[width=0.7\linewidth]{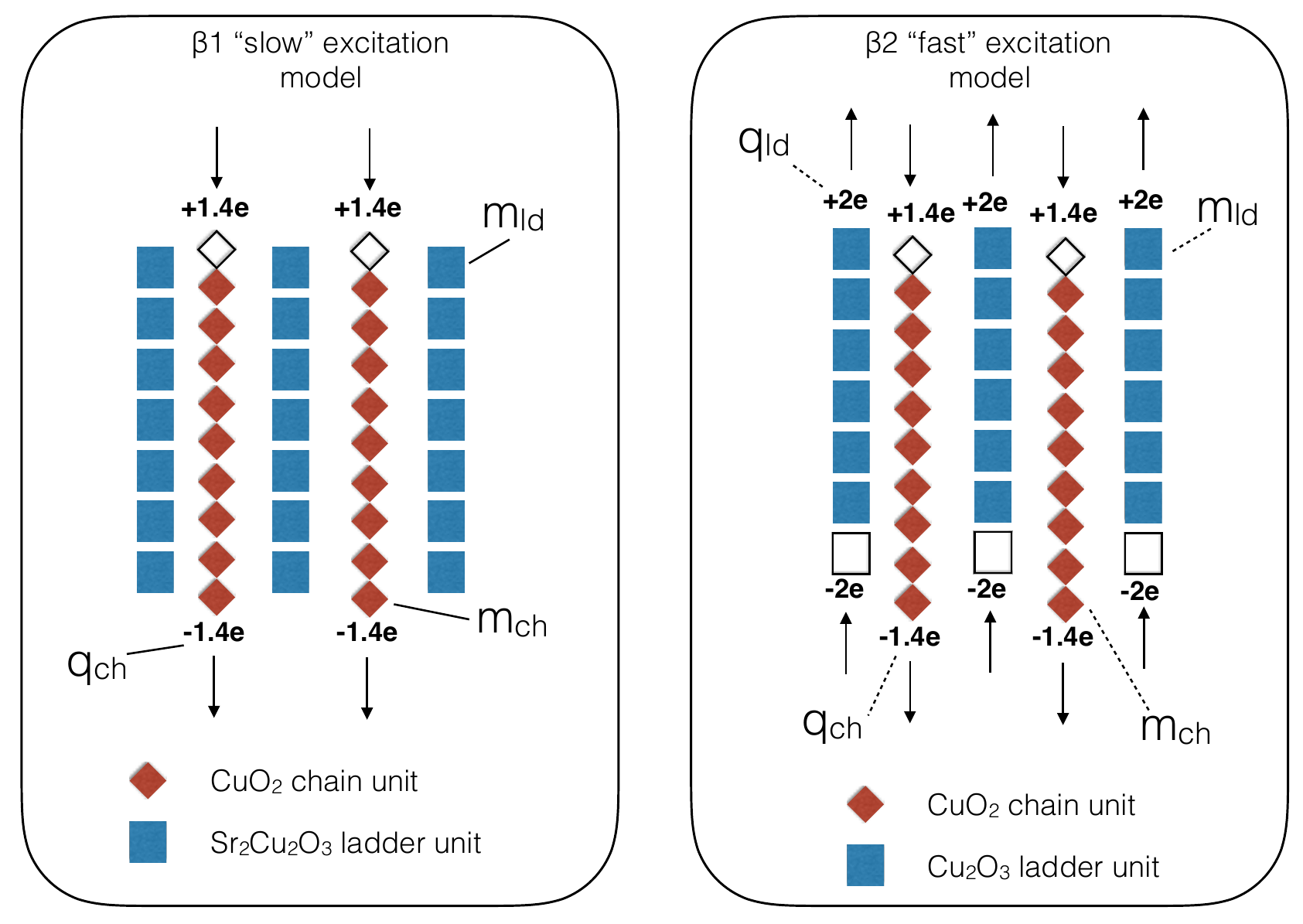}
\caption{Sketch of proposed lattice dynamics for observed excitations $\beta1$ and $\beta2$. The ``slow'' sliding mode ($\beta1$) is explained by the chain (or ladder) sublattice displaced in phase relative to the stationary ladder (or chain) sublattice. The higher order or ``fast'' sliding mode ($\beta2$) is described by an out of phase displacement of each sublattice relative to each other.}
\label{fig:modemodel}
\end{figure}